\documentclass[]{aa}  

\usepackage{graphicx}
\usepackage{txfonts}
\usepackage{xcolor}
\usepackage{amsmath}
\usepackage{amsfonts}
\usepackage{natbib}
\usepackage{longtable}
\usepackage{booktabs}
\usepackage{multirow}
\usepackage[amsmath,thmmarks]{ntheorem}
\usepackage[normalem]{ulem}
\theoremseparator{.}
\theorembodyfont{\normalfont} 
\newtheorem{assumption}{Assumption}{\bfseries}{\itshape}

\defcitealias{2020A&A...644A...7O}{Paper I}

\usepackage[normalem]{ulem}
\usepackage{hyperref}
\hypersetup{
    colorlinks=true,
    linkcolor=blue,
    filecolor=magenta,      
    urlcolor=cyan,
    citecolor=teal,
    pdftitle={Kalkayotl 2.0},
    }
\begin{document}

   \title{Kalkayotl 2.0}

   \subtitle{Bayesian phase-space modelling of star-forming regions, stellar associations, and open clusters}

   \author{J. Olivares\inst{1} \and H. Bouy\inst{2}  \and Trevor Z. Dorn-Wallenstein\inst{3} \and A. Berihuete\inst{4}}

   \institute{
   		Departamento de Inteligencia Artificial, Universidad Nacional de Educación a Distancia (UNED), 
   		c/Juan del Rosal 16, E-28040, Madrid, Spain. 
   		\email{jolivares@dia.uned.es}
		\and
		Laboratoire d'astrophysique de Bordeaux, Univ. Bordeaux, 
   		CNRS, B18N, allée Geoffroy Saint-Hilaire, 33615 Pessac, France.
   		\and
   		The Observatories of the Carnegie Institution for Science, 813 Santa Barbara Street, Pasadena, CA 91101, USA.
   		\and
   		Depto. Estadística e Investigación Operativa, Universidad de Cádiz, Avda. República Saharaui s/n, 11510 Puerto Real, Cádiz, Spain.
        }

   \date{}

 
  \abstract
   {Star-forming regions, stellar associations, and open clusters are fundamental stellar systems where predictions from star-formation theories can be robustly contrasted with observations.}
   {We aim to provide the astrophysical community with a free and open-source code to infer the phase-space (i.e. positions and velocities) parameters of stellar systems with $\lesssim$1000 stars based on \textit{Gaia} astrometry and possibly observed radial velocities.}
   {We upgrade an existing Bayesian hierarchical model and extend it to model 3D (positions) and 6D (positions and velocities) stellar coordinates and system parameters with a flexible variety of statistical models, including a linear velocity field. This velocity field allows for the inference of internal kinematics, including expansion, contraction, and rotation.}
   {We extensively validated our statistical models using realistic simulations that mimic the properties of the \textit{Gaia} Data Release 3. We applied \textit{Kalkayotl} to $\beta$-Pictoris, the Hyades, and Praesepe, recovering parameter values compatible with those from the literature. In particular, we found an expansion age of $19.1\pm1.0$ Myr for $\beta$-Pictoris and rotational signal of $32\!\pm\!11\,\rm{m\,s^{-1}\,pc^{-1}}$ for the Hyades and that Praesepe's rotation reported in the literature comes from its periphery.}
   {The robust and flexible Bayesian hierarchical model that we make publicly available here represents a step forward in the statistical modelling of stellar systems. The products it delivers, such as expansion, contraction, rotation, and velocity dispersions, can be directly contrasted with predictions from star-formation theories.}


   \maketitle
%
\section{Introduction}
\label{introduction}

Open clusters, stellar associations, and star-forming regions, referred to as low-number stellar systems (LNSS), are laboratories where theories of star formation, stellar evolution, and stellar dynamics can be tested and validated. Often, the predictions of these theories are stated in physical spaces rather than in observational ones. Therefore, comparing theoretical predictions to observations requires translating operations that keep biases at the minimum and thoroughly propagate the uncertainty from the observed space to the physical one. The proper handling of this uncertainty propagation is fundamental for hypotheses testing. 

The kinematics of stellar systems are one of their fundamental properties that allows for the testing of theories about their formation, evolution, and disaggregation. Because the kinematic predictions are often stated in the joint space of 3D positions and 3D velocities rather than in the observed space of sky coordinates, parallax, proper motions, and radial velocity, the testing of these theories requires an inference process as a mandatory bridging step.  

The problem of inferring the kinematic parameters\footnote{Through this work, we refer to the parameters of the stellar system as group-level or population-level, parameters, whereas to star-level or source-level parameters to those of individual stars.} of stellar systems  is commonly addressed in the literature from at least three different perspectives. The most common approach is to translate the observations of the ensemble of members to the physical space, and then perform the inference in that space. In contrast, in an alternative approach, the inference is done in the observed space, and then the population parameters are translated to the physical one. Finally, in the third and most computationally expensive approach, the inference of the parameters is done in the physical space by proposing parameter values that are transformed to the observed space, where they are compared to the observations. This last approach is known as forward-modelling because it proposes a model in the space of interest and moves it forward to the observed space, where the likelihood is computed \citep[see, for example,][]{2018A&A...616A...9L}. Unfortunately, none of these perspectives is free of caveats and their application depends on the particular objective of the researcher. We now describe what we consider are the most concerning caveats of these approaches.

The transformation of the data from the observed space to the physical one demands sources with fully observed entries (i.e. radial velocity and all astrometric entries). However, radial velocities are usually only present for a small fraction of the bright stellar system's members. Therefore, the practitioners of this approach tend to discard sources with missing radial velocity despite the quality of their astrometry. The caveat of this approach is that the parameters inferred based on such cropped samples will be biased towards the properties of the brightest observed members and will have larger uncertainties due to a smaller sample. The second approach is usually unaffected by the previous caveat given that the population parameters can be inferred independently in each feature of the observed space, despite missing values. However, transforming the resulting distributions of the  population parameters from the observed space to the physical one requires the Jacobian of the transformation, which, for non-linear transformations, heavily depends on the precision of the measurements.

Moreover, the transformation of either measurements, as done in the first approach, or population parameters, as done in the second one, from the observed space to the physical one, is a non-trivial problem that unavoidably requires prior assumptions. Even in the presence of high signal-to-noise measurements or low-uncertainty parameters, the choice of the prior is of critical importance \citep[see][for the archetypical case of transforming parallaxes into distances]{2015PASP..127..994B}.

The forward-modelling approach is free of the previous caveats but is often time-consuming for both human modellers and computing machines. We now present a brief review of the literature works that use this methodological approach for the task of inferring the internal kinematics of LNSS.

In their pioneering series of articles, \citet{1999A&A...348.1040D} proposed, \citet{2000A&A...356.1119L} developed, and \citet{2002A&A...381..446M} applied a new method to estimate the space velocity and internal velocity dispersion of stellar clusters based on astrometric data alone. This method infers the cluster velocity parameters using a maximum-likelihood formulation together with Monte Carlo simulations. Their basic cluster model includes only the cluster's velocity vector and an internal and isotropic velocity dispersion as population parameters and the parallax of each star as source-level parameters. In their complete model, the entries of a linear-velocity tensor were also inferred as population parameters. One of the advantages of this model is that it produces as output the kinematically improved parallaxes of each star together with its predicted astrometric radial velocity. 

Building upon the previous method, \citet{2020MNRAS.498.1920O} created the open code \textit{Kinesis}, which fits the internal kinematics of open clusters with astrometry and (possibly missing) radial velocity data of its members. The inclusion of radial velocities allowed them to break the degeneracy between expansion or contraction and approaching or receding perspective effects present in the original method by \citet{2000A&A...356.1119L}. Although their method continues to infer the velocity population- and source-level parameters, it relaxes the previous authors' assumption on the isotropy of the velocity dispersion and infers its full covariance matrix. Therefore, their Bayesian forward-model for the velocity field allows for the inference of full velocity dispersion; a linear gradient that incorporates, expansion, rotation, and shear; and a background model for decontamination. 

\citet{2019MNRAS.489.3625C} created the \textit{Chronostar} open code, which uses astrometry and radial velocities (when available) of sources within a specified sky region, to simultaneously identify the LNSS present in the region, deliver a list of their candidate members, infer their phase-space parameters (i.e. positions, velocities, and their dispersion), and estimate kinematic age. However, its application for estimating kinematic age is restricted to young stellar systems \citep[see Sect. 3.4 of][]{2023MNRAS.519.3992Z}. The code infers the stellar system's phase-space parameters of location and dispersion assuming that the dispersion is isotropic but independent in positions and velocities.

\citet{2024MNRAS.527.4193W} developed a Bayesian hierarchical model for studying Galactic globular clusters that allows for the inference of source-level (i.e. heliocentric phase-space coordinates of individual stars) and population-level (i.e. distribution function) parameters, which include, among others, the cluster's position and velocity dispersions. Rather than assuming a classical statistical distribution for the population-level parameters, the authors assume a lower isothermal distribution function. Although designed for globular clusters, the authors validate their method on datasets containing up to a thousand stars. We include this method in our review because it shares similar characteristics to the method presented in this work.

In this work, we aim to provide the astrophysical community with a free and open-source code that simplifies the implementation of forward-models of the phase-space of stellar systems. The methodological and practical implementation (i.e. the source code) that we present here is the multi-dimensional extension of the 1D \textit{Kalkayotl} code \citep[][hereafter \citetalias{2020A&A...644A...7O}]{2020A&A...644A...7O}. Thus we continue using the same name and online address.\footnote{\label{footnote:online_adress}\url{https://github.com/olivares-j/Kalkayotl}} The new version of the code will allow its users to infer the 3D Cartesian or 6D phase-space parameters of LNSS with a variety of flexible models. However, we warn its users that its main caveat is that it remains computationally expensive (see Appendix \ref{appendix:scalability}). 

This work is organised as follows. In Sect. \ref{method}, we introduce the theoretical and practical frameworks to infer the 3D positions and the 6D phase-space parameters of LNSS. Then, in Sect. \ref{validation}, we validate these frameworks with synthetic datasets. In Sect. \ref{application}, we exemplify the use of \textit{Kalakyotl} by applying it to benchmark stellar systems where other literature methods have been applied, particularly, \textit{Chronostar} in $\beta$-Pictoris and \textit{Kinesis} in the Hyades. Finally, in Sect. \ref{conclusions}, we present our conclusions. Appendices \ref{appendix:assumptions} and \ref{appendix:scalability} show a compilation of the method's assumptions and the code's time scalability, respectively.

\section{Methodology}
\label{method}

\textit{Kalkayotl} infers the internal positions and kinematics of LNSS by creating flexible Bayesian hierarchical models with domains in the distance space (i.e. the 1D model), the Cartesian positional space (i.e. the 3D model), or Cartesian phase-space (i.e. the 6D model). These Cartesian coordinates represent source-level parameters from the bottom-level hierarchy of the model, while the population-level parameters represent the upper-layer hierarchy of the model. The space of Cartesian coordinates can be selected to correspond either to the ICRS or Galactic reference systems,\footnote{By specifying the \textit{reference\_system} argument.} for which we followed the \textit{PyGaia}\footnote{\url{https://github.com/agabrown/PyGaia}} conventions and definitions. 

In the rest of this section, we describe each of the dimensionalities of \textit{Kalkayotl}'s models (i.e. 1D, 3D, and 6D) and the families of statistical distributions it provides to describe the data. Throughout this section we undertake the assumptions specified in Appendix \ref{appendix:assumptions}. Later on, we specify our choice of prior distributions for the population-level parameters and the probabilistic programming language that we use to sample the  posterior distributions. We retain the central and non-central parametrisations introduced in \citetalias{2020A&A...644A...7O} and follow the recommendations therein: central parametrisation for distances up to 500 pc and non-central beyond. Finally, we end the section by describing our strategies to diagnose the prior specification, the convergence of the Markov chains, and the predictive power of the inferred models.

\subsection{The 1D distance models}
\label{method:1D_models}
The 1D distance models are described in \citetalias{2020A&A...644A...7O}. Briefly, in those models, we constructed a Bayesian hierarchical model in which the stellar distances (source-level parameters) were sampled from distributions of statistical (e.g. uniform, Gaussian) or astrophyscial (e.g. King's) origin whose (global-level) parameters were inferred hierarchically from the data as well. In this new version of the code, we keep the same 1D models as in \citetalias{2020A&A...644A...7O} and, following \citet{2021AJ....161..147B}, include, as a prior for the distance $r\!>\!0$, the generalised Gamma distribution \citep[GGD;][]{Stacy1962} which is defined as $p(r|a,d,p)=(p/a^d) r^{d-1} e^{-(r/a)^p}/\Gamma_f(d/p)$, with shape parameters $d>0$ and $p>0$, scale parameter $a>0$, and $\Gamma_f$ the gamma function, respectively. Then, Eq. 3 of \citet{2021AJ....161..147B} is recovered, by setting $a=L$, $d=\beta+1$, and $p=\alpha$ as

\begin{equation}
\mathcal{GGD(}r|L,\alpha,\beta)=\frac{1}{\Gamma_f\big(\frac{\beta+1}{\alpha}\big)}\frac{\alpha}{L^{\beta+1}}r^\beta e^{-(r/L)^\alpha}.
\end{equation}
We note that the exponentially decreasing space density (EDSD) prior introduced by \cite[][Eq. 17]{2015PASP..127..994B} is a special case of the $\mathcal{GGD}(r|L,\alpha,\beta)$, and corresponds to $\alpha=1$ and $\beta=2$. When implementing this prior within \textit{Kalkayotl}, we assume the following hyperpriors: for $\alpha$ and $\beta$, we assume uniform distributions with a width of 100, and lower limits set to 0 and -1 respectively. The length scale $L$ is assumed to be Gamma-distributed, with $L\sim\Gamma(\alpha=2,\beta=2/c)$ such that $c>0$ is a user-defined hyper-parameter that corresponds to the mean of the length scale $L$.

\subsection{The 3D position models}
\label{method:3D_models}
We modelled the 3D positions (i.e. the X, Y, and Z source-level parameters) of individual stars (either in ICRS or Galactic coordinates) as independent and identically distributed (\textit{iid}) random variables drawn from the same multivariate parent distribution in the hierarchy, which comprises Gaussian, Student-T, and Gaussian mixture models (GMM). For the GMM, we included two special cases: the concentric GMM (CGMM) and the field GMM (FGMM). In the CGMM all Gaussian distributions in the mixture share the same mean  or location parameter. The FGMM also has concentric mean parameters and a component representing the field in which the covariance matrix is diagonal and fixed to the values provided by the user. Specific details on the number of parameters and their dimensionality is given in the following sections. 

\subsection{The 6D phase-space models}
\label{method:6D_models}
We modelled the 6D coordinates (i.e. X, Y, Z, U, V, W source-level parameters) of individual stars as either joint distributions of positions and velocities or as a linear velocity field. In the linear velocity field, the velocities are expressed as a linear combination of the positions plus a peculiar velocity. As in the 3D models, the user can choose to infer the model parameters in the Cartesian ICRS or Galactic reference systems.

\subsubsection{Joint models}
\label{method:joint_models}
The phase-space joint models are generalisations of the 3D models (see Sect. \ref{method:3D_models}) in which the dimensionality simply goes from 3 to 6, with its consequent increase in the total number of model parameters. For example, the Gaussian and GMM models now have $6\times n + 27$ and $6\times n + 27\times m + m -1$ free parameters, respectively.

\subsubsection{Linear velocity field}
\label{method:linear_models}
In addition to the joint model of positions and velocities described above, \textit{Kalkayotl} allows for the construction of a linear \footnote{By including $\kappa$ and $\omega$ as model parameters.} velocity field model. Following \citet{2000A&A...356.1119L}, we modelled the velocity of a star as the addition of the system's velocity (together with its internal dispersion) plus the velocity expected by a linear field: 
\begin{equation}
\label{equation:linear_model}
\vec{v}_i = \boldsymbol{T}\cdot(\vec{x}_i-\vec{x}_0)+ \mathcal{D}(\vec{v}_0,\Sigma_{\vec{v}}),
\end{equation}
where $\boldsymbol{T}$ is a $3\times3$ tensor representing the linear velocity gradient, $\vec{x}_i$ and $\vec{x}_0$ are the 3D positions of the star and the system, respectively, and $\mathcal{D}(\vec{v}_0,\Sigma_{\vec{v}})$ is either a multivariate Gaussian or multivariate Student-T distribution (see Sects. \ref{method:Gaussian} and \ref{method:Student-T}, respectively) centred at the system's velocity $\vec{v}_0$ and with velocity dispersion given by the covariance matrix $\Sigma_{\vec{v}}$. The entries of tensor $\boldsymbol{T}$ correspond to the nine partial derivatives of the velocity components (U, V, W) with respect to the position components (X, Y, Z). 

Here, we represent the diagonal, lower-triangular, and upper-triangular entries of tensor $\boldsymbol{T}$ with vectors $\boldsymbol{\kappa}$, $\boldsymbol{\Omega}_0$, and $\boldsymbol{\Omega }_1$, respectively, as follows:

\begin{equation}
\label{equation:linear_tensor}
\boldsymbol{T}=\left(\begin{array}{ccc}
T_{xx} & T_{xy} & T_{xz} \\ 
T_{yx} & T_{yy} & T_{yz} \\ 
T_{zx} & T_{zy} & T_{zz}
\end{array} \right)=\left(\begin{array}{ccc}
\kappa_x & \Omega_{00} & \Omega_{01} \\ 
\Omega_{10} & \kappa_y & \Omega_{02} \\ 
\Omega_{11} & \Omega_{12} & \kappa_z
\end{array} \right).
\end{equation}

With this notation, the mean of vector $\vec{\kappa}$ expresses the internal contraction or expansion of the system, with $|\vec{\kappa}|=\frac{1}{3}(T_{xx}+T_{yy}+T_{zz})=\frac{1}{3}(\kappa_{0}+\kappa_{1}+\kappa_{2})$ while the rotational velocity vector $\vec{\omega}$ can be expressed as: $\vec{\omega}\,=\,[\omega_x,\omega_y,\omega_z]\,=\,\frac{1}{2}[T_{zy}-T_{yz},T_{xz}-T_{zx},T_{yx}-T_{xy}]$ \citep[see Eq. 4 of][]{2000A&A...356.1119L}. We notice that a simpler and constant\footnote{By including only $\kappa$ and not $\omega$ as model parameter.} velocity model can be obtained by neglecting the contributions of vectors $\boldsymbol{\Omega}_0$ and $\boldsymbol{\Omega}_1$ which effectively reduce tensor $\boldsymbol{T}$ to a diagonal matrix. 

Thanks to the previous definitions, objective criteria can be defined to establish if a stellar system is expanding, contracting or rotating. We use the following definitions throughout the rest of this work. Expansion or contraction is said to be detected at the $\alpha$ level if the $\alpha$-quantile of the high-density interval (HDI) from the posterior distribution of $|\vec{\kappa}|$ does not contain zero. For example, expansion (contraction) is said to be detected at the $2\sigma$ level if the 95\% HDI of the posterior distribution of $|\vec{\kappa}|$ is positive (negative) and does not contain zero. Similarly, we define a detectability criterion for rotation. We said to detect rotation at the $\alpha$ level if the $\alpha$-quantile HDI of the posterior distribution of at least one of the entries of the $\vec{\omega}$ vector does not contain zero. We notice that our restrictive criterion applies to each independent component given that a set of rotation vectors with random orientations may also have a magnitude that could be significantly positive without implying an ordered motion. In the case of expansion or contraction, given that the important quantity, $|\vec{\kappa}|$, is defined as the average of the trace of tensor $T$ and there are only two directions (contraction or expansion) there is no issue in defining its significance over the posterior distribution of $|\vec{\kappa}|$ rather than on the components of $\vec{\kappa}$. 

The previous velocity model is essentially the same that \citet{2000A&A...356.1119L} originally proposed and then \citet{2020MNRAS.498.1920O} augmented by a background modelling. There are nonetheless the following differences. First, our model lacks an integrated decontamination mechanism (see Assumption \ref{assumption:clean_members}), as the iterative outliers rejection of \citet{2000A&A...356.1119L} or the background modelling of \citet{2020MNRAS.498.1920O}. Second, the 3D positions in our model are inferred using a system-oriented hierarchical model (see Sect. \ref{method:families}), which offers more robust estimates than the improper 1D distance prior assumed by \citet{2020MNRAS.498.1920O}. Third, observed radial velocities are also included as part of the data; thus the perspective expansion or contraction effects affecting the \citet{2000A&A...356.1119L} model are no longer present. As stated in Appendix A of \citet{2020MNRAS.498.1920O}, a few radial velocity measurements across the system serve to anchor its systemic velocity and remove these perspective effects. Finally, the major difference between the previous models and ours is the correction of the angular (spatial) correlations present in both proper motions and parallaxes of \textit{Hipparcos} \citep{1997A&A...323L..49P} and \textit{Gaia} \citep{2016A&A...595A...1G} data. Correcting for these correlations results in improved parameter precision and credibility (see \citetalias{2020A&A...644A...7O}).

Regarding contamination, we believe that due to the multitude of origins that it may have (e.g. observation conditions, reduction pipeline, catalogue creation, selection function, membership methods, cherry-pick selection) a single decontamination model will not be optimal for a particular research objective given that an object may be considered a contaminant in some cases but not in others. The clearest example of this type of objects is an unresolved binary star, which could be considered a contaminant for some objectives (e.g. trace-back age dating) and not in others (e.g. mass function determination). For the previous reasons, we decided to leave the decontamination process to the criteria of the user. Nonetheless, we provide a simple decontamination method similar to that of \citet{2020MNRAS.498.1920O} but constructed in the joint space of positions and velocities, which can be run independently from the rest of the models (see Sect. \ref{method:FGMM}). The user can access this model through the \textit{FGMM} family and the specific \textit{field\_scale} parameter. The probabilistic classification resulting from this model can be used to clean the list of system members.

\subsection{Families of statistical distributions}
\label{method:families}

In \citetalias{2020A&A...644A...7O}, we use two types of families to describe the distance distributions of stellar systems: the purely statistical and the astrophysical ones. Here, we only provide purely statistical distributions and leave the inclusion of astrophysical ones for future work. The purely statistical distributions that we implemented here are Gaussian, Student-T, and GMM, together with two specific cases of the GMM: the concentric Gaussian mixture model and the field Gaussian mixture model.

We notice that in all our model families, their parameters can be held fixed to a user-provided value during the inference process. This proved particularly useful in those cases where the user wants to reproduce the inference process given all or some parameter values from the literature or from a previous run. For example, if the user is interested in inferring the parameters of a single star that is known to be a member of the stellar system for which its parameters are known. Thus, instead of running a full analysis with the previous members plus the new one, the user can assume that this new member has a negligible contribution to the population-level parameters and hold them fixed when inferring its source-level parameters, which will certainly speed up the inference process.  

\subsubsection{Gaussian distribution}
\label{method:Gaussian}

The Gaussian distribution describes the multivariate coordinates (i.e. 3D or 6D) $\vec{x}$ of each star as $\vec{x}\sim\mathcal{N}(\vec{\mu},\Sigma)$, with $\mathcal{N}$ the multivariate normal distribution, $\vec{\mu}$ its vector of central location (i.e. mean or median), and $\Sigma$ the symmetric positive semi-definite covariance matrix. The number of free parameters of the Gaussian distribution equals $D + (D\times(D+1))/2$, with $D$ the model's dimension (i.e. 3 or 6). 

\subsubsection{Student-T distribution}
\label{method:Student-T}
Similar to the Gaussian distribution, the Student-T distribution describes the multivariate coordinates (i.e. 3D or 6D) $\vec{x}$ of each star as $\vec{x}\sim\mathcal{T}(\vec{\mu},\nu,\Sigma)$, with $\vec{\mu}$ and $\Sigma$ defined as in the Gaussian distribution, but now, the $\nu$ parameter describes the degrees of freedom. We notice that when $\nu\rightarrow\infty$ the Student-T distribution converges to the Gaussian distribution. Its number of free parameters is only one more than those of the Gaussian distribution with the same dimensionality. When this family distribution is used in combination with the linear velocity model, the $\nu$ parameter  is a vector of dimension two with the first one used for  positions and the second one for velocities.

We decided to include the Student-T distribution because it describes variations from normality using a single parameter ($\nu$) with clear interpretability and simplicity. The smaller the value of the $\nu$ parameter, the larger the discrepancy with respect to the Gaussian distribution. Its heavy tails for small values of $\nu$ allow for a simple and robust modelling of possible outliers while maintaining a symmetric distribution.

\subsubsection{Gaussian mixture}
\label{method:GMM}
The Gaussian mixture distribution describes the multivariate coordinates (i.e. 3D or 6D) $\vec{x}$ of each star as drawn from a mixture of Gaussian distribution with a fixed number of components $m$ set by the user, that is 
\begin{equation}
\vec{x}\sim\mathcal{GMM}\left(\{w_i,\vec{\mu}_i,\Sigma_i\}_{i=1}^{m}\right)\equiv\sum_{i=1}^m w_i\cdot\mathcal{N}(\vec{\mu}_i,\Sigma_i), \nonumber
\end{equation}
with $w_i$, $\vec{\mu}_i$ and $\Sigma_i$ the weight, mean, and covariance matrix of the i-th Gaussian component, where $\sum_{i=1}^m w_i = 1$. The total number of free parameters for the GMM distribution equals $m\times[(2\times D) + (D\times(D+1))/2] -1 $, with $D$ the model's dimensionality.

We decided to include this family distribution to model complex stellar regions composed of more than one structure. We notice that each component in the mixture has its own location and dispersion, thus representing the most flexible of our models. In addition, this family allowed for a probabilistic posterior classification of individual stars into the components of the mixture, thus enabling probabilistic disentanglement of stellar populations (assuming that the Gaussian distribution is a good model for each of these populations).   

\subsubsection{Concentric Gaussian mixture}
\label{method:CGMM}
The concentric Gaussian mixture (CGMM) as its name says, is a special case of GMM in which the Gaussian components are concentric, this is, they all share the same and unique mean $\vec{\mu}$. Fixing this unique mean effectively reduces the number of free parameters to $D+m\times[D + (D\times(D+1))/2] -1$, with $D$ the model's dimensionality. 

We included this special case of the GMM to describe complex distributions of stellar systems having tails heavier than those of the Student-T distribution but that are symmetric with respect to the centre (i.e. they are concentric). This distribution proved to be useful for describing the long tidal tails of open clusters \citep[see, for example,][]{2023A&A...675A..28O}.

\subsubsection{Field Gaussian mixture}
\label{method:FGMM}
The field plus Gaussian mixture (FGMM) distribution is a further special case of the CGMM in which the last component of the mixture has a fixed and user-provided diagonal covariance matrix. This last component is used to model possible field contaminants whose large dispersion may be known a priori by the user. Therefore, the 6D \textit{field\_scale} vector parameter that represents the standard deviations of the field component should be supplied by the user based on existing a prior information about possible contaminants. The number of free parameters of this model is thus reduced to $D+(m-1)\times[D + (D\times(D+1))/2]-1$, with $D$ the model's dimensionality.

We included this special case to allow for the decontamination of the input list of members. Given the probabilistic classification provided by Gaussian mixtures, the user can make a first run using this decontamination model, and then use its output classification to remove from the input list of members those sources classified as field stars. An example of this procedure is shown in Sect. \ref{application}.

\subsection{Prior specification}
\label{method:prior}

All the family distributions implemented in \textit{Kalkayotl} share two types of parameters: locations and scales. Other parameters, such as the weights in the mixture distributions or the degrees of freedom in the Student-T distribution, are family specific and are discussed below. 

The location parameter corresponds to the single $D$-dimensional vector $\vec{\mu}$ in the Gaussian, Student-T, CGMM, and FGMM families, and to the set of $m$ $D$-dimensional vectors $\{\vec{\mu}\}_{i=1}^{m}$ in the GMM family. Similarly, the scale parameter corresponds to the single $DxD$ covariance matrix $\Sigma$ in the Gaussian and Student-T families and to the set of $m$ $DxD$ covariance matrices $\{\Sigma_i\}_{i=1}^{m}$ in the mixture models ($m-1$ in the case of the FGMM).

For the location parameter ($\vec{\mu}$ or $\{\vec{\mu}\}_{i=1}^{m}$) we used a Gaussian prior. In each entry (coordinate), $\mu$, of this vector parameter we imposed a univariate Gaussian prior so that $\mu_i\sim \mathcal{N}(\mu\!=\!\alpha_{i,0},\sigma = \alpha_{i,1})$, with fixed hyper-parameters $\alpha_{i,0}$ and $\alpha_{i,1}$, corresponding to the median and standard deviation of the Gaussian distribution. The values of these hyper-parameters are either provided by the user or set to its default values.

For the scale parameter ($\Sigma$ or $\{\Sigma_i\}_{i=1}^{m}$) we used independent prior distributions over the correlation matrices, $Corr$, and the diagonal standard-deviation matrix, $Sd$, and reconstructed the scale parameters as $\Sigma = Sd\cdot Corr \cdot Sd$. We imposed a \citet{LEWANDOWSKI20091989} distribution (hereafter LKJ) as prior for the correlation matrices, $Corr \sim \mathcal{LKJ}(\eta)$, with fixed hyper-parameter $\eta$, which controls the degree of correlation between entries (coordinates), with $\eta=1$ resulting in uniform correlations (i.e. $\rho\sim\mathcal{U}(0,1)$) and increasingly larger values producing decreasing correlations. The value of this $\eta$ parameter can be provided by the user or left to its default value of one. In each entry of the diagonal standard-deviation matrix we imposed a Gamma prior, $\sigma_i \sim \Gamma(\alpha=2,\beta=\beta_i)$ with fixed hyper-parameter $\beta_i$ either to a user-provided value or to its default of 10 pc in positions and 2 $\rm{km\,s^{-1}}$ in velocities, which result in modes at 10 pc and 2 $\rm{km\,s^{-1}}$, respectively. The combination of these two hyper-parameters can help users to flexibly infer the shape of the stellar system in both positions and velocities or, on the contrary, to impose spherical symmetry on it.

For the weight parameter $\vec{w}$ of the mixture distributions (i.e. GMM, CGMM and FGMM), we used a Dirichlet prior distribution so that $\vec{w}\sim \mathcal{D}(\vec{\delta})$, with $\vec{\delta}$ a fixed hyper-parameter whose value is provided by the user. This hyper-parameter controls the amount of mass distribution going to each component, with small magnitude vectors resulting in more relaxed priors. For the degrees-of-freedom parameter, $\nu$, of the Student-T distribution we use a Gamma prior distribution so that $\nu \sim \Gamma(\alpha\!=\!\nu_\alpha,\beta\!=\!\nu_\beta)$, with hyper-parameters $\nu_\alpha$ and $\nu_\beta$ fixed to either the user-provided values or $\nu_\alpha=1$ and $\nu_\beta=0.1$.

For the parameters $\vec{\kappa}$ and $\vec{\Omega}$ of the linear velocity field, we use the following prior distributions. The components of both, $\vec{\kappa}$ and $\vec{\Omega}$ are expected to be small and centred on zero. Thus we used zero-centred normal distributions with standard deviations as user-defined hyper-parameters $\sigma_{\vec{\kappa}}$ and $\sigma_{\vec{\Omega}}$, both having a default value of $0.1\rm{km\,s^{-1}\,pc^{-1}}$.

In Table \ref{table:priors}, we provide a summary of the shared and family-specific parameters together with their prior distributions. The right column of the table also provides the default hyper-parameter values for cases in which the user does not provide input values. In the default values of the hyper-parameters, the $\overline{X}$ vector corresponds to the mean vector of observables after transformation into the ICRS or Galactic reference system.

\begin{table}[ht!]
\caption{Prior distributions and its hyper-parameters.}
\label{table:priors}
\centering
\resizebox{\columnwidth}{!}{
\begin{tabular}{ccc}
Parameter & Prior & Default values \\
\hline
\hline 
Location: $\mu_i$ & $\mathcal{N}(\mu\!=\!\alpha_{i,0},\,\sigma\!=\!\alpha_{i,1})$ &  $\alpha_{i,0}\!=\!\overline{X}_i$, $\alpha_1\!=\!0.2\overline{X}_i$\\
Scale: $\sigma_i$ & $\Gamma(\alpha\!=\!2,\beta\!=\! \beta_i)$ &  $\beta_{i\in[X,Y,Z]}\!=\!10$ pc, $\beta_{i\in[U,V,W]}\!=\! 2\, \rm{km\cdot s^{-1}}$\\
Scale: $Corr$ & $\mathcal{LKJ}(\eta)$ &  $\eta\!=\!1$\\ 
Weights: $\vec{w}$ & $\mathcal{D}(\vec{\delta})$ & • \\ 
Degrees of freedom: $\nu$ & $\Gamma(\alpha\!=\!\nu_\alpha,\beta\!=\!\nu_\beta)$ & $\nu_\alpha\!=\!1$, $\nu_\beta\!=\!10$ \\ 
Expansion/Contraction: $\boldsymbol{\kappa}$ & $\mathcal{N}(\mu\!=\!0,\,\sigma\!=\!\sigma_{\kappa})$ & $\sigma_\kappa=0.1\rm{km\,s^{-1}\,pc^{-1}}$\\
Rotation: $\boldsymbol{\Omega}$ & $\mathcal{N}(\mu\!=\!0,\,\sigma\!=\!\sigma_{\Omega})$ & $\sigma_\Omega=0.1\rm{km\,s^{-1}\,pc^{-1}}$\\
\hline
\end{tabular}
}
\tablefoot{{\scriptsize The $i$ index goes over the coordinates $X,\, Y,\, Z,\, U,\, V$ and $W$.}}
\end{table}

\subsection{Likelihood}
\label{method:likelihood}
We assumed that the likelihood of the $N$ stars in the input list of the system's members is a multivariate Gaussian distribution. The dimensionality of this Gaussian corresponds to the total number of effectively observed features in the input list of members $N_{eff}$. For example, the likelihood of a dataset with $N=100$ stars with fully observed astrometry and radial velocity will be a multivariate Gaussian of $N_{eff}=600$ dimensions in the 6D model and $N_{eff}=300$ dimensions in the 3D model. However, the likelihood  of a dataset with the same number of stars ($N=100$) but with only 50 of them having radial velocity measurements would be a multivariate Gaussian with $N_{eff}=550$ and $N_{eff}=300$ dimensions in the 6D and 3D models, respectively. This single high dimensionality Gaussian results from dropping the usual assumption that observations are independent and identically distributed. Instead, we assume that the data is heteroscedastic and spatially correlated in the sky, as is the case of the \textit{Hipparcos} \citep[e.g][]{1988ashc.rept..179L,2000A&A...356.1119L} and \textit{Gaia} \citep[e.g.][]{2021A&A...649A...2L} data. We model these spatial (angular) correlations using Equations 24 and 25 of \citet{2021A&A...649A...2L} (Assumption \ref{assumption:angular_correlations}).

In the previous multivariate Gaussian likelihood, the median value corresponds to the flattened vector of the measurements of all the members in the input list. The covariance matrix of this Gaussian likelihood is constructed as follows. First, the covariance matrix of each source is reconstructed from its provided uncertainties (i.e. the \texttt{\_error} suffix) and correlations (i.e. the \texttt{\_corr} suffix). These $N$ covariances are used to populate the block diagonal entries of the $N_{eff}\times N_{eff}$ covariance matrix. The inter-source covariances (off-block entries) of the parallax and proper motions features are populated with the values provided by the spatial correlation functions mentioned above (see Assumption \ref{assumption:angular_correlations}). We notice that the entries of this covariance matrix corresponding to radial velocities and sky coordinates have zero inter-source correlations.

We noticed that the tiny sky uncertainties (typically of the order of 50 $\rm{\mu\!as}$) of the \textit{Gaia} data difficult the sampling of the model parameters. In a Markov chain, the probability of rejecting a proposed step increases with a diminished likelihood value. Therefore, the tiny uncertainties result in Markov chains with very few accepted transitions, large autocorrelations, and low effective sample sizes. The solution to this problem is to run longer chains at the cost of a higher computing time. We provide an alternative and more practical solution which is to increase the sky uncertainties to a value that results in an acceptable compromise between computing time and convergence of the Markov chains. We heuristically find that increasing the sky uncertainties by factors of the order of $10^5 - 10^6$ (which effectively results in sky uncertainties of the order of 5-50 arcseconds) improves the convergence of the chains with negligible impact on the recovered parameter values. We set the default value of this scaling factor to $10^6$, which renders the best compromise between computing time and convergence assurance.

\subsection{Posterior sampling}
\label{method:sampling}

We specify our models and sample their posterior distributions given the input list of members using the probabilistic programming language \textit{PyMC} \citep{Salvatier2016}. To sample the posterior distributions, \textit{Kalkayotl} finds suitable initial solutions for each Markov chain through auto-differentiable variational inference \citep[ADVI;][]{JMLR:v18:16-107}. This fast initial solution serves to reduce the computation time that otherwise the chains would need to reach the target region of the parameter space. These initial solutions (one per chain) are passed to the No-U-Turn sampler \citep[NUTS;][]{2011arXiv1111.4246H}, which is used for efficient sampling of high-dimensionality spaces. Further sampling acceleration can be achieved using graphical processor units (GPUs) through the high-performance computing library JAX\footnote{\url{http://github.com/google/jax}}. 

The sampler default parameters were established as follows. The ADVI search for the initial solution runs for $5\times 10^5$ steps or an absolute tolerance of $5\times 10^{-3}$. The sampling is done by running two chains with 3000 and 2000 iterations each for tuning and sampling, respectively. The default parameters for NUTS are a target acceptance of 0.65 \citep[see ][]{2011arXiv1111.4246H} and step sizes of $10^{-1}, 10^{-2}$ and $10^{-3}$ for models of 1D, 3D, and 6D, respectively.  These values were found after extensive validation and they probed to effectively reduce the tuning iterations without compromising the convergence. Nonetheless, the 3000 tuning iterations are used to refine this value at each run.

\subsubsection{Convergence and statistics}
\label{method:convergence}

Convergence of the NUTS chains is automatically assessed by \textit{PyMC} at the end of each run. We recommend users to carefully read the output messages of the sampler, which generally provide useful diagnose. Nonetheless, \textit{Kalkayotl} computes its convergence assessment through the following statistics. The potential scale reduction \citep{1992StaSc...7..457G}, known as the Gelman-Rubin $\hat{R}$ statistic, is intended to measure the factor by which the posterior would shrink if the number of iterations will go to infinity. It is intended for multiple chains and its value should be $\lesssim 1$, with values of $\hat{R}>1.1$ indicating convergence issues. The effective sample size (ESS) measures the number of independent samples with the same statistical power as the $N$ autocorrelated samples from the Markov chain \citep[see Sect. 11.5 of ][]{gelmanBDA}, and thus the Markov chain standard error (MCSE) of the parameters is $\sigma/\sqrt{\rm{ESS}}$ rather than $\sigma/\sqrt{N}$, with $\sigma$ the standard deviation. These three statistics are reported for each model parameter as part of the standard output files. In addition, the terminal output of \textit{Kalkayotl} also reports the step size value of the NUTS algorithm used by each chain, which can be used to further refine its value in consecutive runs and to diagnose convergence issues when different chains provide highly discrepant step size values. 

\subsubsection{Model criticism: Prior and posterior checks}
\label{method:criticism}

Model criticism is a fundamental step of the knowledge extraction process. Although it lies beyond the inference process itself, it is important to diagnose possible biases associated with the model itself rather than with the sampling algorithm. For this reason, model criticism must be done once the convergence of the sampling algorithm has been warranted. 

The most straightforward element to criticise in a Bayesian parametric model is its set of prior distributions. This criticism of the prior in most cases can be done by visual comparison of the prior and posterior distributions. Upon user's requests,  \textit{Kalkayotl} samples from the prior predictive distribution and thus generates samples from the prior distribution of the model parameters. Then, the code provides output files with plots of these prior distributions together with the inferred posterior ones. Inspecting these plots, the user can easily spot problematic prior distributions. In principle, the prior distribution should allow, with a non-negligible probability, the expected intervals of the parameter value without being overly restrictive to avoid biasing the posterior distribution towards specific parameter regions.

The second and sometimes most difficult element to criticise in a model is its set of prior assumptions (i.e. the assumptions that the modeller made when constructing the model). Assessing the fitness of these assumptions is dataset specific and far from simple. One diagnostic that aids in this criticism, particularly when comparing different models, is the ability of the model and its inferred parameter distributions to predict observed data, which could be sources observed or unobserved by the model. \textit{Kalkayotl} provides, as part of the output files, the posterior predictive distributions of the model's observed sources (i.e. those used to infer the posterior distributions). These synthetic observed data can be used to judge the model's ability to predict the real observed data. And thus, by comparing the predictive posterior distributions of different models, the user can decide the specific model that better suits the particular needs of the scientific objective at hand. 

We notice that these posterior predictive distributions may have further astrophysical uses. For example, by comparing the radial velocities posterior predictive distributions with the observed ones, the user can discriminate between single stars and possible binaries or multiples. Also, when working with 6D models, the posterior predictive distributions of the parallaxes would in general shrink with respect to the observed ones due to the effect of the kinematically improved parallaxes \citep[see][and its Sect. 4.1.2, particularly]{2000A&A...356.1119L}. One further use could be to obtain kinematic parallaxes by adding to the system's input dataset the data of candidate members with missing parallaxes but measured sky positions and proper motions (in the same reference system as \textit{Gaia}), as those provided by the DANCe \citep{2013A&A...554A.101B} and VISIONS \citep{2023A&A...673A..58M} surveys.

\section{Validation}
\label{validation}

We validated our methodology using synthetic data sets with \textit{Gaia} data release 3 \citep[DR3,][]{2023A&A...674A...1G} properties created with the \textit{Amasijo}\footnote{\url{https://github.com/olivares-j/Amasijo}} code. For detailed explanations of this code, we refer the reader to the works of \citet{2022A&A...664A..31C} and \citet{2023A&A...675A..28O}. Briefly, the code creates a synthetic population of stars from the user-provided kinematic parameters (location and dispersion in the ICRS or Galactic reference systems, optionally a linear velocity tensor can also be provided) together with the fixed population's age, metallicity, and extinction. These true values are transformed into the observed space (astrometry plus photometry) and then resampled from the observational uncertainties given by \textit{PyGaia}\footnote{\url{https://github.com/agabrown/PyGaia}} for \textit{Gaia} DR3. The radial velocities are masked as missing according to \textit{PyGaia} criteria, which roughly correspond to sources fainter than 14 mag in G band. Therefore, the percentage of sources with missing radial velocities varies from 10\% for clusters at 100 pc to 50\% for those at 1.5 kpc. 

For each model family, we created synthetic clusters in a grid with varying numbers of stars $n\_stars\!\in\![100, 200,400]$, distances $d\!\in\![50, 100, 200,400,800,1500]$ pc, and random seeds $s\!\in\![0,1,2,3,4,5]$. We fixed the population's age, metallicity and extinction to 100 Myr, $z=0.012$, and $A_v=0.0$, respectively. The system's location in space and velocity space was fixed to $XYZ=[1,1,1]\cdot d/\sqrt{3}\,\rm{pc}$ and $UVW=[10,10,10]\,\rm{km\,s^{-1}}$ whilst its scale (dispersion) fixed to 3 pc in position and 1 $\rm{km\,s^{-1}}$ in velocity, both isotropic with zero correlations.

We generated stellar systems with five statistical models: Gaussian joint, Gaussian linear, Student-T joint, Student-T linear, and GMM joint. The joint models are multivariate in the joint space of positions and velocities ($XYZUVW$) while the linear ones are disjoint in positions ($XYZ$) and velocities ($UVW$). In the linear velocity field, the velocities are expressed as a linear field of the positions (see Eq. \ref{equation:linear_model}) with a tensor $T$ fixed to

\begin{equation}
\label{equation:tensor_values}
\boldsymbol{T}=C\cdot\left(\begin{array}{rrr}
 1 & -1 &  1 \\ 
 1 &  1 & -1 \\ 
-1 &  1 &  1
\end{array} \right)\, \rm{m\, s^{-1}\,  pc^{-1}}
\end{equation}

with $C$ as a constant taking values in $C\in[10,50,100]$. Such tensor produces stellar systems that expand isotropically at a rate $|\kappa|\!=\! C\,\rm{m\, s^{-1}\, pc^{-1}}$ and rotate with an angular velocity $\vec{\omega}\!=\![C,C,C]\,\rm{m\, s^{-1}\, pc^{-1}}$. The chosen values of $C$ are similar to the expansion and rotation signals found in the literature for the benchmark stellar systems of the $\beta$-Pictoris stellar association and the Hyades and Praesepe open clusters (see Sect. \ref{application}). In the Student-T and GMM, we fixed the parameter values to $\nu\!=\!10$ and $w\!=\![0.6,0.4]$ with two components, respectively. We notice that in all cases we use the same location and scale parameters as described above, although in the GMM case, the second component is always located 50 pc above the first component in the Z direction.

We inferred the stellar system parameter of each synthetic data set using its corresponding model family. The posterior distribution was sampled, analysed, and criticised as described in Sects. \ref{method:sampling}, \ref{method:convergence}, and \ref{method:criticism}, respectively. Appendix \ref{appendix:scalability} shows some examples of the typical execution time that \textit{Kalkyotl} takes as a function of the number of stars in the system. 

In the following, we validate and assess the quality of the recovered parameters with three metrics: error, uncertainty, and credibility. The error measures the parameter's relative deviation from its true value and is computed as the difference between the mean of the parameter's posterior distribution and the true value divided by the true value. The uncertainty metric corresponds to the standard deviation of the parameter's posterior distribution divided by the parameter's true value. These error and uncertainty metrics correspond to the ones typically known as accuracy and precision, although here we express them in relative terms. Finally, the credibility metric measures the fraction of the simulations in which the true parameter's value was contained within the 95\% HDI of the parameter's inferred posterior distribution. All of these metrics are expressed in percentage units.

In the next two sections, we present the validation of the population-level parameters and the source-level parameters. Finally, we end this section with a discussion about our models'  characteristics.

\subsection{Population-level parameters}
\label{validation:group-level}

We validate the population-level parameters of our set of models with the error, uncertainty, and credibility metrics mentioned above. First, we assess the performance of our method to recover the set of parameters that are common to all models (i.e. the location and standard deviation). Then, we continue the validation and discussion of the model-specific parameters, particularly those from the linear velocity field.

\subsubsection{Common parameters}
\begin{figure*}[h!]
    \centering
     \includegraphics[width=0.85\textwidth,page=1]{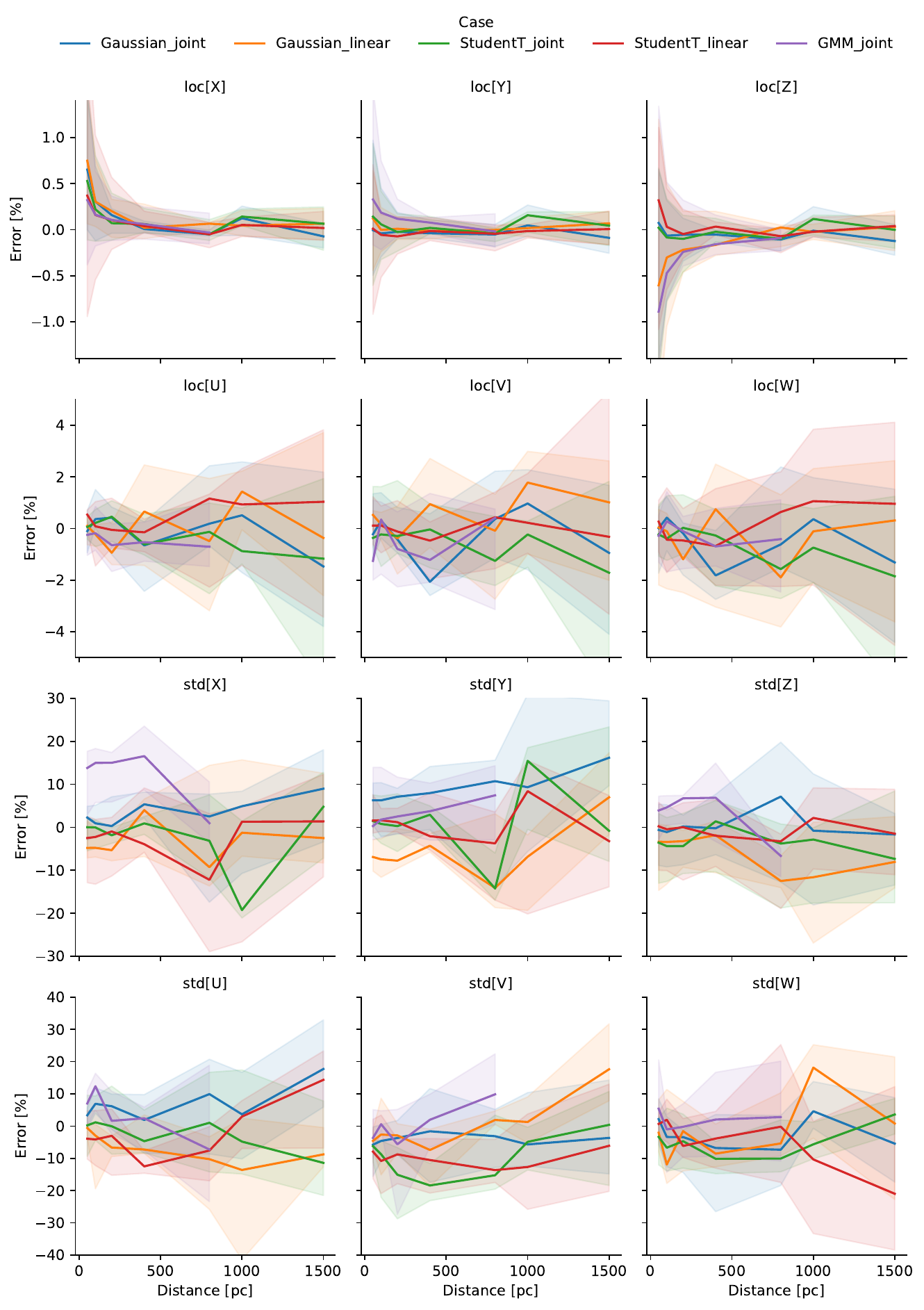}
     \caption{Relative error of the population-level parameters common to all our models as a function of distance. The lines and shaded regions depict the mean and standard deviation of the synthetic clusters with 100 stars.}
\label{figure:common_parameters_err}
\end{figure*}

\begin{figure*}[h!]
    \centering
     \includegraphics[width=0.85\textwidth,page=2]{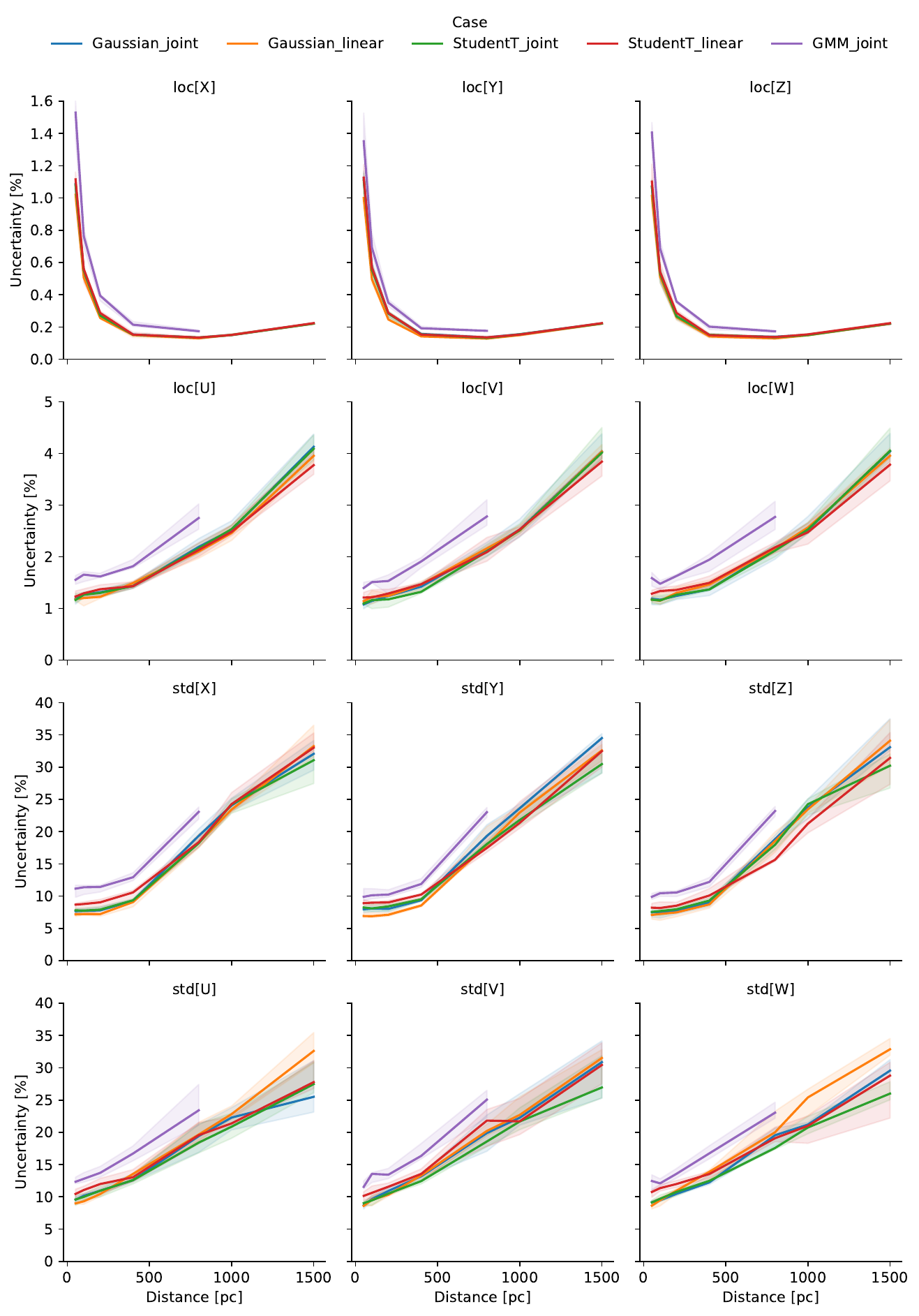}
     \caption{Relative uncertainty of the population-level parameters common to all our models as a function of distance. Captions as in Fig. \ref{figure:common_parameters_err}.}
\label{figure:common_parameters_unc}
\end{figure*}

Figures \ref{figure:common_parameters_err} and \ref{figure:common_parameters_unc} show the error and uncertainty, respectively, of the population-level parameters of location and standard deviation, which are common to all our models, as a function of the cluster's distance. The solid lines and shaded regions depict, respectively, the mean and standard deviation of the random cluster realisations at each distance value. The figures show only the results on synthetic clusters with 100 stars given that the metrics of clusters with 200 and 400 stars are better, as expected, and thus, for the sake of simplicity, we do not show them. In the case of the GMM model, we only show the result of one component and up to a distance of 800 pc. Beyond this limit, the parameter's recovery shows large errors, large uncertainties and negligible credibility. From these figures, we draw the following conclusions.

First, the errors and uncertainties are isotropic showing similar values in the position and in the velocity coordinates. The isotropy of our results indicate that our models can recover the true phase-space geometry of stellar systems with errors $\lesssim$10\% up to 800 pc and $\lesssim$20\% up to 1.5 kpc, which implies that up to 800 pc there is a minimal "fingers of God" effect \citep[for a description of this effect, see, for example, Fig. 2 of][]{2020A&A...633A..51Z}. The only exception to this trend is the GMM model, which shows larger $std$ errors, particularly in the X and Z directions. These large errors most likely result from the combination of both the lower number of stars (the shown GMM component has only 60 stars compared to the rest of the cases having 100 stars each) and the entanglement between components in the Z direction resulting from the second component being displaced 50 pc in this direction.

Second, concerning the errors, we observe that all our models, except for the GMM one, have negligible errors in the location parameters of position ($\lesssim$0.5\%) and velocity ($\lesssim$3\%), with the former having larger values in the closest clusters and the latter in the farthest ones. The errors in the standard deviation parameters of all models are $\lesssim$20\% for both positions and velocities, with their values increasing from 5-10\% at the closest clusters and up to 20\% for the farthest ones. The GMM model shows the largest dispersion error in the X and Z directions for the reasons explained in the previous paragraph.

Third, concerning uncertainties, the 3D location of the clusters are recovered with uncertainties in the 0.5-1.5\% range for clusters up to 200 pc and $\lesssim$0.2\% for clusters up to 1.5 kpc. The relatively larger uncertainties at the closest distances indicate that our models struggle to infer the 3D centre of the nearest clusters despite the excellent precision and accuracy with which the 3D positions of their stars are recovered (at the 0.1\% level, see Figs. \ref{figure:source-level_err} and \ref{figure:source-level_unc}). In the rest of the parameters and models, the uncertainties grow with distance, as expected, from 1\% to 5\% in the velocity location, and from 10\% to 35\% in the rest of the dispersion parameters.

Concerning the parameter's credibility, the median value of this metric for all location and standard deviation parameters in all models is 100\%, except for the GMM one beyond 800 pc, where it drops to zero as mentioned above. The high credibility indicates that our method can recover the true parameters values within the 95\% HDI for distances up to 800 pc for the GMM model and up to 1.5 kpc for the rest of the models.

Finally, the low error, low uncertainty, and high credibility prove that our method is able to recover the true parameters of location and standard deviation of stellar systems with excellent accuracy and acceptable precision in the GMM model up to 800 pc and up to 1.5 kpc in the rest of models. Most probably, the subset of the location parameters could still be recovered with excellent accuracy and precision beyond 1.5 kpc. However, we are interested in showing the validity of the model as a whole rather than subsets of it.

\subsubsection{Model-specific parameters}

There are three sets of additional model-specific parameters: the $\nu$ parameter of the Student-T distribution, the weight parameters of the GMM model, and the nine entries of the $T$ tensor in the linear velocity field models. In the following, we briefly discuss those of the GMM and Student-T, and then we discuss in more detail the parameters of linear velocity field models, particularly their detectability.

The $\nu$ parameter in the Student-T models is recovered with large errors for the closest clusters (200\% in the joint model and 60\% and 40\% in the position and velocity of the linear field model) but these errors diminish with distance and stabilise at 800 pc with values of 100\% for the joint model and <5\% for the linear field model. This parameter is recovered with large uncertainties ($\sim$100\%) in both the joint and linear field models for all distances. In spite of the large errors, the credibility reaches 100\% due to the large uncertainties, except in the case of the joint model at closer distances ($\lesssim$100 pc), where the credibility is  lower than 60\% due to the large errors. Therefore, we recommend caution when using the Student-T model and either disregarding the inferred values of the $\nu$ parameter or fixing it to a certain a priori value before the inference.

The weights of the GMM model are recovered with negligible errors, $\lesssim$2\%, low uncertainties, $\lesssim$12\%, and 100\% credibility for distances up to 800 pc. Beyond this limit, the errors and uncertainties increase and the credibility goes to zero. Therefore, we conclude that our method can accurately and precisely recover the true weights of the GMM model to distances up to 800 pc.

The entries of the linear velocity tensor $T$ (see Eq. \ref{equation:linear_tensor}) are all recovered with almost identical metrics. Thus, the following description is valid for any of the entries.

Concerning errors, these vary mostly with the $C$ value and are almost independent of the number of stars. For the lowest value of $C\!=\!10$, the errors are independent of distance and their absolute values are, on average, $\lesssim$200\%.  For $C\!=\!50$, the errors remain independent of distance in general, but this time, their absolute values are $\lesssim$50\%. For the largest value of $C\!=\!100$, the errors show a linear trend with distance that varies between 0\% at 50 pc to -50\% at 1.5 kpc.

Concerning the uncertainties, these grow with distance and diminish with an increasing number of stars, as expected. For $C\!=\!10$, they vary between 200\% at 50 pc to 800\% at 1.5 kpc, while for $C\!=\!50$ they vary between 40\% at 50 pc to 160\% at 1.5 kpc. Finally, for $C\!=\!100$, the uncertainties vary between 20\% at 50 pc to 80\% at 1.5 kpc. Despite the large errors, the large uncertainties result in a high 100\% credibility for all $C$ values, distances and number of stars.

In the light of the previous results, we recommend proceeding with caution whenever an entry of the linear velocity tensor is inferred with a value $\lesssim\!\!50\,\rm{m\,s^{-1}\,pc^{-1}}$, particularly when it is lower than $\lesssim\!\!10\,\rm{m\,s^{-1}\,pc^{-1}}$. The criteria for the detection of such low signals should be based on their full posterior distribution rather than only on their mean value.

\begin{figure}[ht!]
    \centering
     \includegraphics[width=\columnwidth]{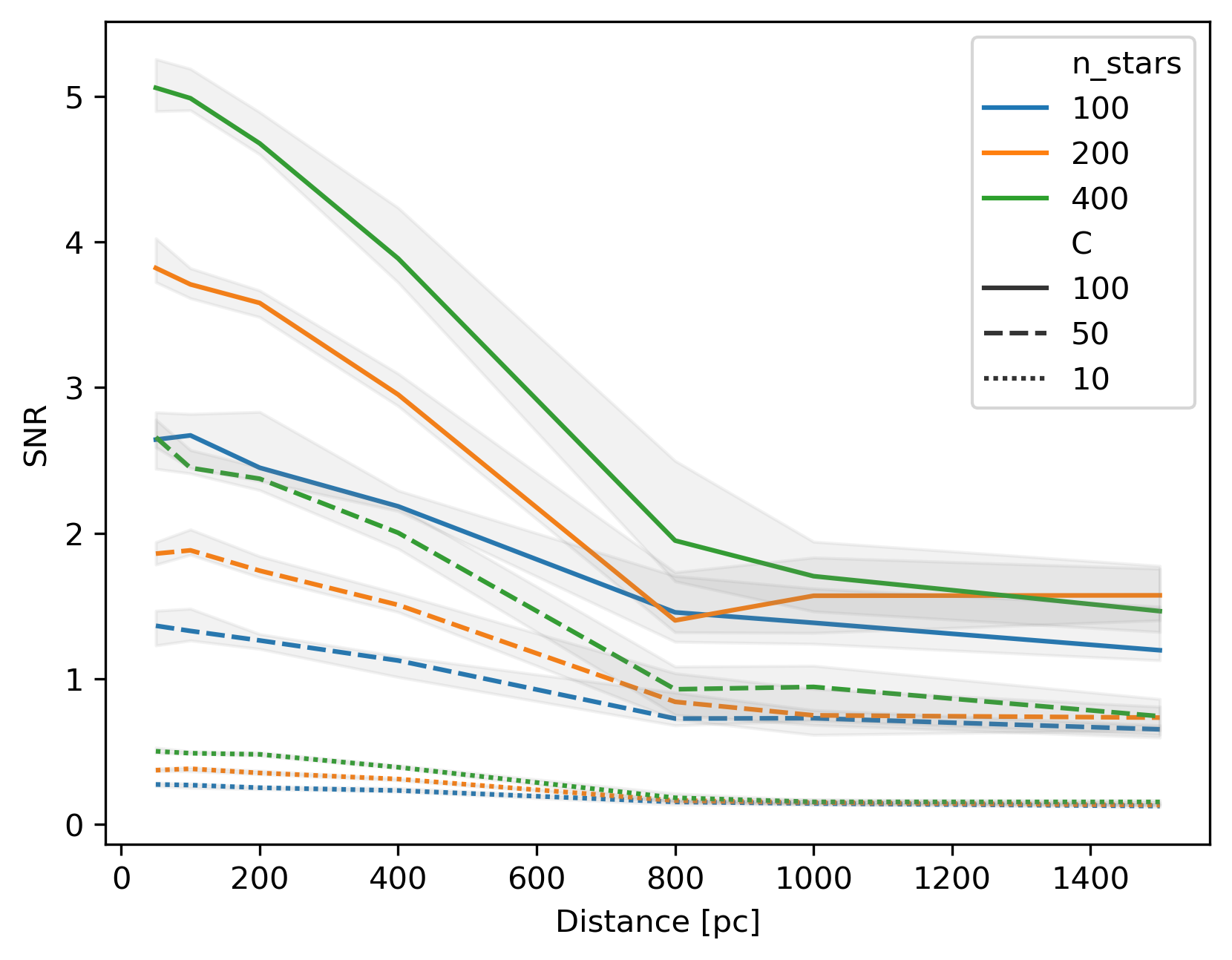}
     \caption{Signal-to-noise ratio of the entries in the linear velocity tensor as a function of the distance and the number of stars for the three values of the $C$ constant in units of metre per second per parsec (see Eq. \ref{equation:tensor_values}). The lines and shaded regions depict the median and standard deviation of the synthetic simulations with varying random seed. }
\label{figure:snr}
\end{figure}

Usually, the detection of a signal is based on its significance level, with 3$\sigma$ to 5$\sigma$ being common criteria to establish a discovery. Moreover, the significance level of a signal can be related to the well-known term of signal-to-noise ratio (S/N), in which an S/N=$n$ is roughly equivalent to a $n\sigma$ significance level.

To aid the users of \textit{Kalkyotl} to estimate the detectability of the entries in the linear velocity tensor, we use our set of \textit{Gaia} DR3 simulations to compute the S/N of the inferred $T$ tensor entries. In Fig. \ref{figure:snr}, we show the S/N of these entries as a function of the cluster distance, and colour-coded with the number of stars for the three values of the $C$ constant (shown with different line style). As can be observed, the S/N improves with signal value and number of stars and diminishes with distance up to a maximum distance of 800 pc, beyond this limit, it remains almost constant independent of the number of stars. Moreover, the figure shows that given the current \textit{Gaia} DR3 data (i.e. astrometry and radial velocity) a 5$\sigma$ detection can only be achieved in the nearest (50 pc) and most populated clusters (>400 stars). However, if the criteria would be relaxed to 1$\sigma$ significance level, then the detection of signals $\sim\!50\,\rm{m\,s^{-1}\,pc^{-1}}$ could be achieved up to distances of 800 pc even for clusters with 100 stars. We noticed that these S/N improve with the improving quality of the radial velocities. Thus, complementing \textit{Gaia} DR3 astrometry with precise radial velocity surveys could significantly improve the detection of low $T$ entry signals.

\subsection{Source-level parameters}
\label{validation:source-level}

\begin{figure*}[ht!]
    \centering
     \includegraphics[width=0.9\textwidth,page=1]{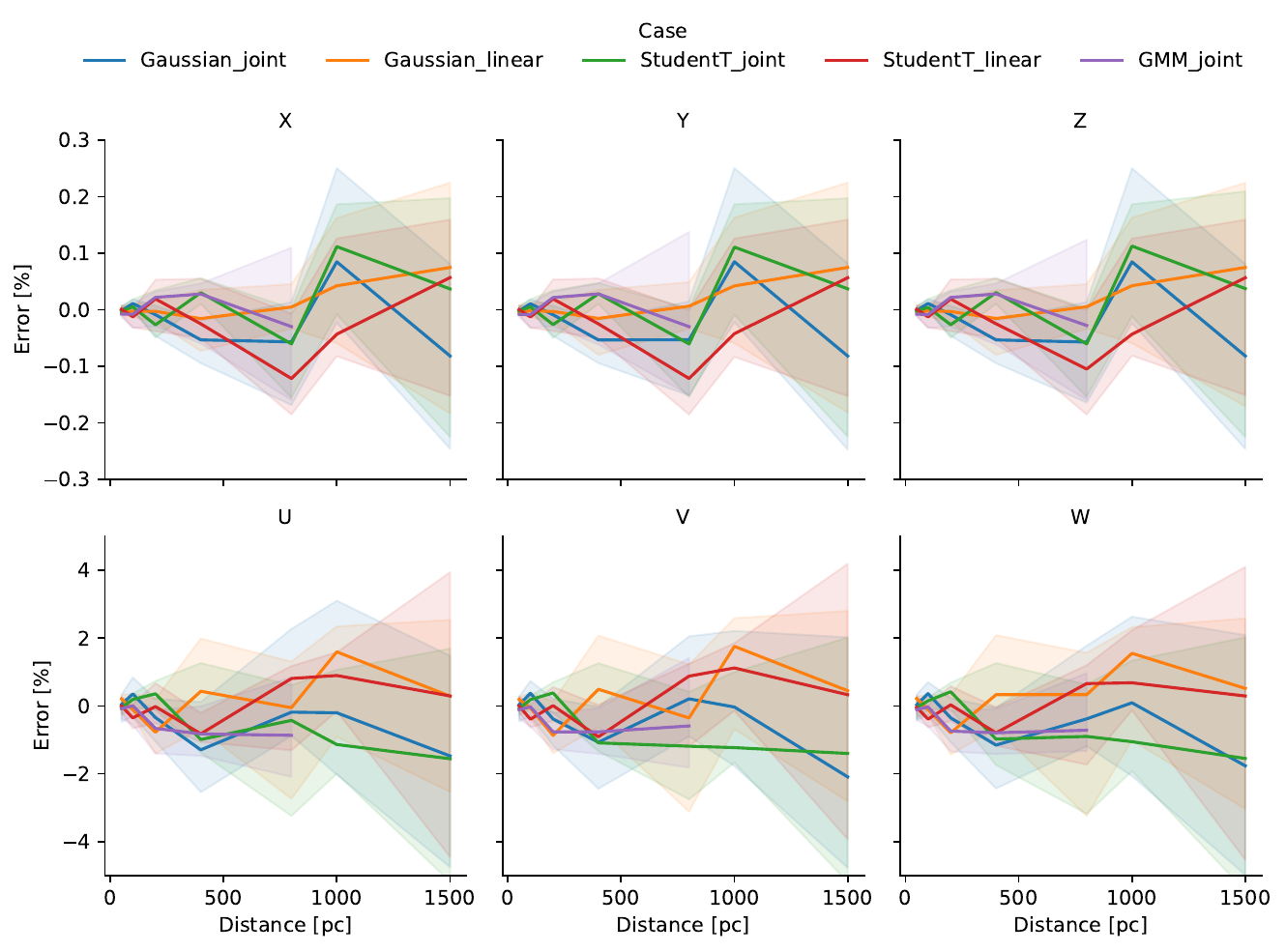}
     \caption{Relative error of the source-level parameters as a function of distance. Captions as in Fig. \ref{figure:common_parameters_err}.}
\label{figure:source-level_err}
\end{figure*}

\begin{figure*}[ht!]
    \centering
     \includegraphics[width=0.9\textwidth,page=2]{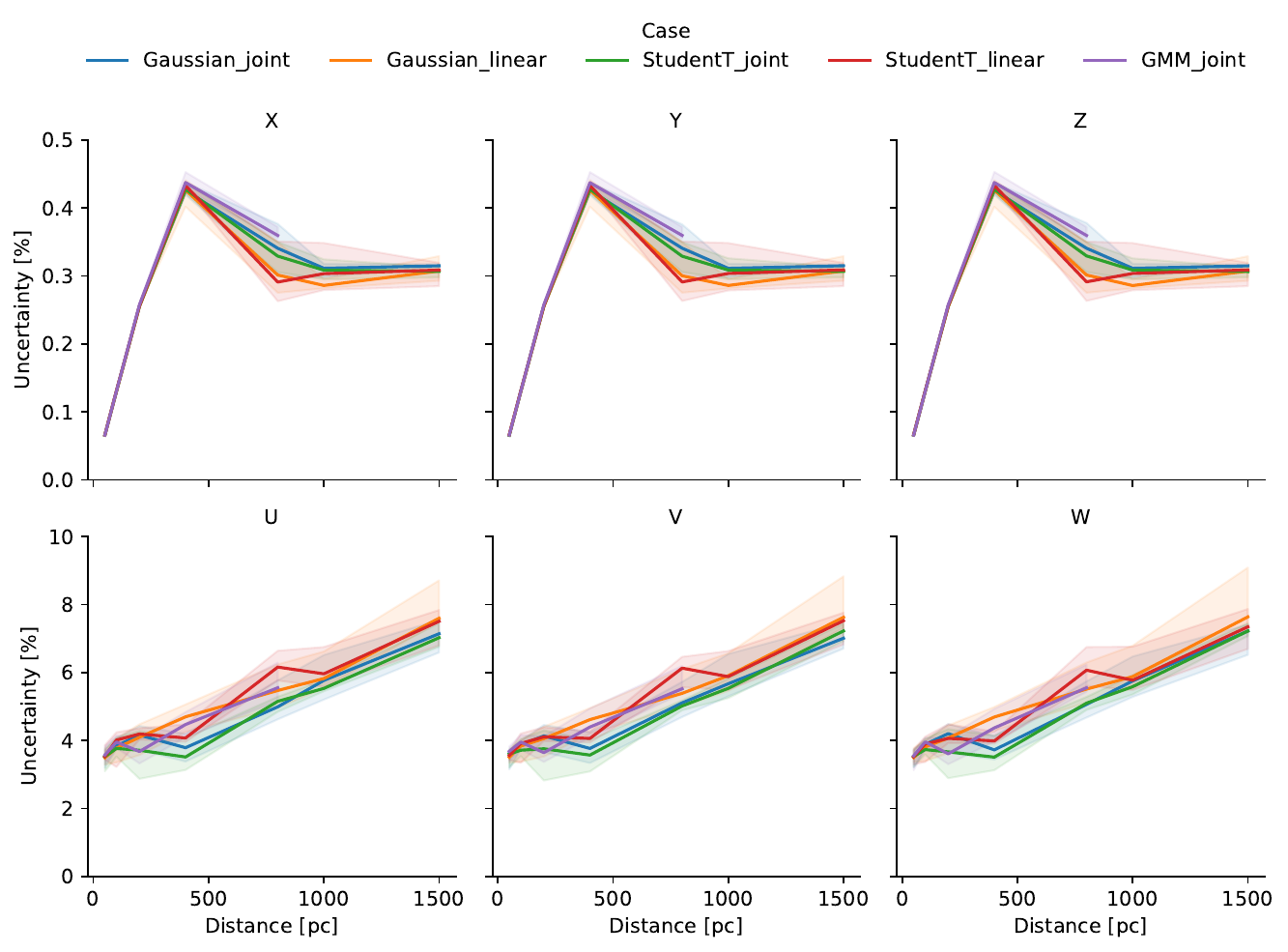}
     \caption{Relative uncertainty of the source-level parameters. Captions as in Fig. \ref{figure:common_parameters_err}.}
\label{figure:source-level_unc}
\end{figure*}

We validate the source-level parameters (i.e. the phase-space coordinates of individual stars) with the same metrics used for the population-level parameters. However, in this case, we first averaged over the entire system's population and then over simulations. The error and uncertainty in the source-level parameters of the six phase-space coordinates are shown in Figs. \ref{figure:source-level_err} and \ref{figure:source-level_unc}, respectively. The figures show the metrics value as a function of distance colour-coded according to the model. As with the figures of the group-level parameters, the ones here do not show the values of the GMM model for distances beyond 800 pc nor the results of clusters having 200 and 400 stars. The metrics of clusters having larger populations remain similar to those of having 100 stars except for a mild error reduction in the clusters at 1.5 kpc. In the case of the GMM model, as discussed in the previous section, the maximum distance at which it can be reliably applied is 800 pc. Thus, we do not show values beyond this limit. In the rest of the distances and models, the credibility remains high at 100\%, and thus, for the sake of simplicity, their figures are not shown. In the following, we discuss the properties of the previous metrics.

As in the case of the population-level parameters, the source-level ones do not show signs of the "fingers of God" effect. The errors and uncertainties are isotropic with similar values for the 3D positions and 3D velocities.

Concerning errors, we observe that the X, Y, and Z coordinates are recovered with excellent accuracy, having errors $\lesssim$0.1\%. Although the U, V, and W velocities are recovered with larger errors, $\lesssim$2\%, these are still negligible.

The uncertainty of the source-level parameters is also remarkable, with values $\lesssim$0.4\% for 3D positions and $\lesssim$8\% for 3D velocities. These later behave as expected, growing at a constant rate of 4\% at 50 pc to 8\% at 1.5 kpc. On the contrary, the uncertainty of the 3D positions shows two distinct behaviours. First, it grows with a high slope from values $\lesssim$0.1\% at 50 pc to 0.45\% at 400 pc. Then, it decreases to 0.3\% at 800 pc and remains constant at this value. These two behaviours result from the use of the central and non-central parametrisations (see Sect. \ref{method}), in which the former is used for clusters up to 500 pc and the latter for farther away ones. As can be seen in Fig. \ref{figure:source-level_unc}, the use of these parametrisations ensures that the uncertainty remains minimal.

In \citetalias{2020A&A...644A...7O}, we observed that the distance errors of the individual sources showed an anti-correlation with their true position within the cluster (see Sect. 4.2 and Fig. 2 of \citetalias{2020A&A...644A...7O}). In our phase-space models, this anti-correlation remains, although much less pronounced. For the 3D positions, it is negligible at 50 pc, grows up to -0.5 at 500 pc and remains at this value for larger distances. On the contrary, for the 3D velocities, it remains at -0.5 for all distances. The explanation for this anti-correlation remains the same as in the 1D case. It results from the ratio between the size of the source uncertainty to that of the cluster, coordinate by coordinate. When the source's uncertainty is much smaller than the cluster's dispersion, then there is enough information to pinpoint the source position within the cluster and the anti-correlation is negligible, as in the case of the 3D positions at 50 pc. As the source's uncertainty increases and gets similar or larger than the cluster's dispersion, then there is not enough information to pinpoint the source's position within the cluster and its posterior gets attracted towards the cluster's mode, thus  resulting in the anti-correlation.

Finally, we can conclude that despite the previously described anti-correlations, the low errors, small uncertainties, and high credibility prove that our method can recover the phase-space coordinate of stars in LNSS with excellent accuracy and precision.

\subsection{Discussion}
\label{validation:discussion}

The results of the previous sections show that \textit{Kalkayotl} is able to recover the true values of the population-level and source-level parameters from various phase-space family distributions and velocity models with varying degrees of accuracy and precision. Despite the remarkable similarities in the metrics values, we notice the following important differences amongst models.

First, the GMM model has a lower applicability domain, with a recommended maximum distance of 800 pc. This reduced domain results from increasing entanglement of the GMM Gaussian components with increasing dataset uncertainties. We recommend that the users of this model do a thorough exploration of the chain's traces to spot possible label switching. Moreover, when the uncertainties are large and the population is scarce, the model can fail to detect one of the components. For this reason, we recommend running the code with several chains or more than only one time. In addition, this model faces convergence difficulties beyond 500 pc. Whenever this occurs, we recommend increasing the number of iterations in the initialisation and tuning phases.

Second, the Student-T family, with both joint and linear velocity fields, shows good metrics values in the common parameters of location and standard deviation. However, the low credibility of the $\nu$ parameter must be kept in mind when using this model.

Third, due to its excellent accuracy and precision, our recommended model is the Gaussian one. In those cases where the use of the GMM model is unavoidable, we recommend nonetheless performing a run of the Gaussian model on each of the identified GMM components.

Finally, we observe that when the linear velocity field is applied to clusters at close distances or with a large number of sources, the sampling algorithm faces difficulties in the convergence of the chains. These difficulties result from both the model's large number of parameters (nine more than the joint model) and its lack of flexibility (given that velocities are assumed to follow Eq. \ref{equation:linear_model}). Although these issues can easily be solved by increasing the number of iterations in the initialisation and tuning phases, we recommend informing the model on the system's velocity dispersion by setting its prior to sensible values, for example, to those obtained after running the joint velocity model on the same system.

\section{Application to real stellar systems}
\label{application}
As part of the validation process and to exemplify the use of \textit{Kalkayotl} in real data, we use it to infer the 6D phase-space source- and population-level parameters of the \object{$\beta$-Pictoris} stellar association and the \object{Hyades} and \object{Praesepe} open clusters. We select these systems because they have ample literature works describing their internal kinematics and relatively low ($\lesssim 1000$) numbers of members, which keeps the computation time at a reasonable value (see Appendix \ref{appendix:scalability}). In $\beta$-Pictoris we use both the Gaussian joint and Gaussian linear velocity models, whereas in the open clusters we only use the Gaussian linear velocity one. Other examples of the application of \textit{Kalkayotl}'s Gaussian, GMM, and CGMM family distributions with the joint velocity model to open clusters and star-forming regions can be found in \citet{2023A&A...671A...1O} and \citet{2023A&A...675A..28O}. The users of \textit{Kalkayotl} can find all the associated routines to infer and analyse the following stellar systems in the same online address as the source code (see footnote \ref{footnote:online_adress}, specifically the folder \texttt{article/v2.0/Code}).

\subsection{The $\beta$-Pictoris stellar association}
\label{beta_pic}

\begin{table*}[ht!]
\caption{Location and standard deviation parameters of the $\beta$-Pic stellar association as reported in the literature and inferred in this work with the Gaussian joint model.}
\label{table:beta_pic_grp}
\centering
\resizebox{0.8\textwidth}{!}
{
\begin{tabular}{llrrrrrr}
\toprule
 &  & \multicolumn{2}{c}{Couture+2023} & \multicolumn{2}{c}{Crundall+2019} & \multicolumn{2}{c}{Miret-Roig+2020} \\
 &  & Reported & This work & Reported & This work & Reported & This work \\
Parameter & Units &  &  &  &  &  &  \\
\midrule
loc[X] & [pc] & $22.69$ & $23.90\pm4.59$ & $30.00\pm3.15$ & $25.90\pm3.44$ & $47.49\pm0.11$ & $38.35\pm4.29$ \\
loc[Y] & [pc] & $-4.31$ & $-2.90\pm2.86$ & $-5.50\pm2.80$ & $-4.81\pm1.96$ & $-7.89\pm0.04$ & $-10.65\pm2.73$ \\
loc[Z] & [pc] & $-18.49$ & $-18.97\pm1.77$ & $-17.50\pm1.70$ & $-18.16\pm1.58$ & $-17.92\pm0.05$ & $-17.12\pm1.77$ \\
loc[U] & [km/s] & $-10.20$ & $-10.13\pm0.26$ & $-9.60\pm0.20$ & $-9.59\pm0.24$ & $-8.74\pm0.24$ & $-8.77\pm0.27$ \\
loc[V] & [km/s] & $-15.70$ & $-15.58\pm0.12$ & $-15.74\pm0.10$ & $-15.63\pm0.14$ & $-16.16\pm0.11$ & $-15.93\pm0.14$ \\
loc[W] & [km/s] & $-8.64$ & $-8.67\pm0.16$ & $-8.85\pm0.10$ & $-8.72\pm0.15$ & $-9.98\pm0.11$ & $-9.03\pm0.15$ \\
std[X] & [pc] & $29.70$ & $29.56\pm4.01$ & $24.50\pm1.45$ & $27.76\pm2.89$ & $16.04$ & $25.00\pm3.46$ \\
std[Y] & [pc] & $13.94$ & $15.84\pm2.44$ & $21.60\pm1.15$ & $14.02\pm1.58$ & $13.18$ & $13.84\pm2.07$ \\
std[Z] & [pc] & $8.11$ & $9.45\pm1.52$ & $13.70\pm0.85$ & $10.78\pm1.18$ & $7.44$ & $8.95\pm1.41$ \\
std[U] & [km/s] & $1.50$ & $1.53\pm0.23$ & $1.20\pm0.10$ & $1.73\pm0.20$ & $1.49$ & $1.51\pm0.21$ \\
std[V] & [km/s] & $0.60$ & $0.59\pm0.10$ & $0.90\pm0.10$ & $0.90\pm0.12$ & $0.54$ & $0.69\pm0.11$ \\
std[W] & [km/s] & $0.76$ & $0.75\pm0.12$ & $1.00\pm0.10$ & $0.95\pm0.12$ & $0.70$ & $0.75\pm0.12$ \\
\bottomrule
\end{tabular}
}
\end{table*}

Young stellar associations are one of the fundamental products of star formation. As such, studying their kinematics is paramount to understanding their origin and evolution. The $\beta$-Pictoris ($\beta$-Pic) stellar association, due to its proximity and young age (28 pc and 18-20 Myr for dynamical ages \citealt{2023ApJ...946....6C,2020AA...642A.179M,2019MNRAS.489.3625C}, $24\pm3$ Myr from isochrones \citealt{2015MNRAS.454..593B} and 24-25 Myr from Lithium depletion boundary \citealt{2022A&A...664A..70G,2016A&A...596A..29M}),  has been the focus of several studies and, as a result, its kinematic properties are well constrained. Here, we use the \textit{Gaia} DR3 data of the lists of members from \citet{2023ApJ...946....6C}, \citet{2020AA...642A.179M}, and \citet{2019MNRAS.489.3625C}. In the case of \citet{2019MNRAS.489.3625C}, we use the \textit{Gaia} DR3 data of their 46 members in component A with a membership probability greater than 0.9. In the cases of \citet{2020AA...642A.179M} and \citet{2023ApJ...946....6C}, we use the \textit{Gaia} DR3 data of their 26 and 25 members, respectively, complemented by the radial velocities provided in each work. 

The previous works report the parameter values in the Galactic reference systems, and thus, for comparison, we used it as well. We notice though that \citet{2019MNRAS.489.3625C} report their parameters with respect to a spatial origin located 25 pc above the Galactic plane and a velocity origin coinciding with the local standard of rest\footnote{There is a typo in the sign of their $V_\odot$ velocity component.}. We subtracted this origin to put their values in the same reference frame as the other two works. We also notice that \citet{2023ApJ...946....6C} do not report uncertainties to any of their parameters, while \citet{2020AA...642A.179M} only report those of the location parameters. 

Table \ref{table:beta_pic_grp} shows the group-level parameters (i.e. location and standard deviation together with their $\pm\sigma$ uncertainties) of the $\beta$-Pic stellar association as reported in the aforecited literature works \citep[i.e.][]{2023ApJ...946....6C,2020AA...642A.179M,2019MNRAS.489.3625C} and inferred here using \textit{Kalkayotl}'s 6D Gaussian joint model. For each of the selected literature works, the table shows the reported parameter's values and those inferred here with the \textit{Gaia} DR3 data of the corresponding authors' membership list.

As can be observed, there is a general agreement between the values inferred here and those reported in the literature works, with the majority being compatible within the 2$\sigma$ uncertainty (i.e. the 95\% HDI). However, the largest exceptions correspond to the X parameters (both location and standard deviation) reported by \citet{2020AA...642A.179M}, and those corresponding to the standard deviation in Y, Z, and U as reported by \citet{2019MNRAS.489.3625C}. Given the large X location and low X dispersion reported by \citet{2020AA...642A.179M} as compared to the rest of the estimates, we consider that these large deviations are probably an artefact resulting from either the membership list or the authors' method. Similarly, the discrepant standard deviations values reported by \citet{2019MNRAS.489.3625C} in the Y, Z, and U coordinates are most likely the result of their method, given that in the same membership list as those authors we find parameters' values perfectly compatible with those reported by \citet{2023ApJ...946....6C} and \citet{2020AA...642A.179M}. Therefore, we conclude that our parameters estimates are generally compatible with those from the literature at the 2$\sigma$ level.

Concerning our parameter's uncertainty, we observe that these are generally similar to those reported by the literature works. In particular, there is a remarkable similarity between our uncertainties and those reported by \citet{2019MNRAS.489.3625C}. With respect to the parameters reported by \citet{2020AA...642A.179M}, we observe that in the location velocities both works have similar uncertainties. However, we notice that their uncertainties in the X, Y, and Z locations are smaller by one and two orders of magnitude. Given that our uncertainty values are similar to those of other parameters and literature works, we conclude that those reported by \citet{2020AA...642A.179M} in the location of the X, Y, and Z coordinates are most likely underestimated.

It is interesting to highlight that, in spite of the different samples of members used by the previous works, our methodology renders all the inferred parameters ($loc$ and $std$) compatible within the $2\sigma$ (95\%HDI) uncertainties, except in the case of $loc[U]$ were the values inferred from the samples of \citet{2020AA...642A.179M} and \citet{2023ApJ...946....6C} are mutually exclusive.

\begin{figure*}[ht!]
    \centering
     \includegraphics[width=\textwidth,page=1]{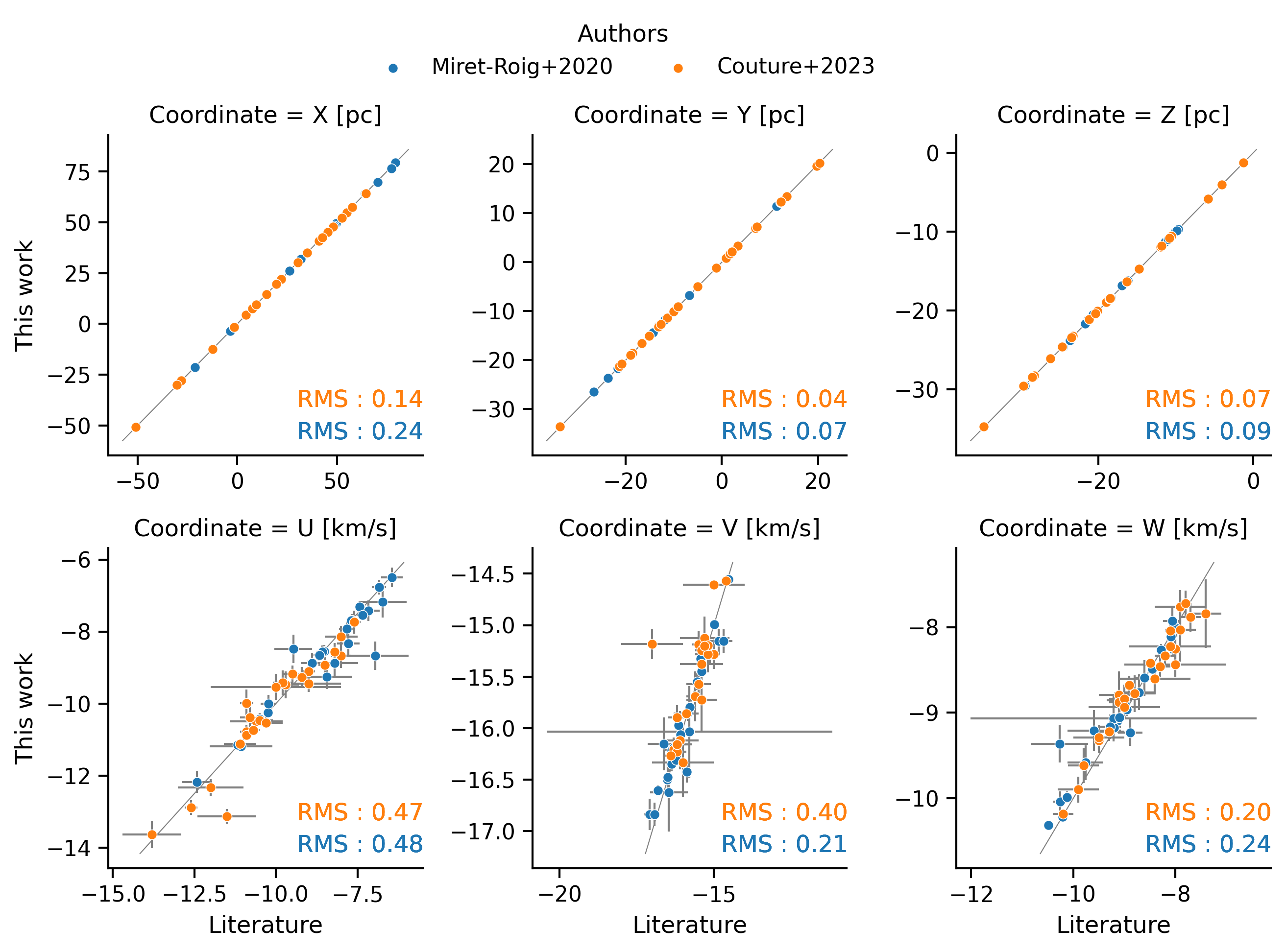}
     \caption{One-to-one comparison of the source-level parameters inferred here and those reported by \citet{2023ApJ...946....6C} and \citet{2020AA...642A.179M}. The grey line depicts the identity relation. The root-mean-squared value of the differences is also shown on the lower right side of each panel.}
\label{figure:beta_pic_src}
\end{figure*}

In Fig. \ref{figure:beta_pic_src}, we show a one-to-one comparison of the source-level parameters (i.e. the coordinate values of each star) as inferred here against those reported in \citet{2023ApJ...946....6C} and \citet{2020AA...642A.179M}. As can be observed, we recover with excellent accuracy and precision the values reported by those authors. The X, Y, and Z, coordinates have an average error $\lesssim$0.1-0.2 pc while  in the U, V, and W it is $\lesssim0.5\,\rm{km\, s^{-1}}$.

\begin{table*}[ht!]
\caption{Expansion, rotation, and age of $\beta$-Pic.}
\label{table:betapic}
\centering
\resizebox{\textwidth}{!}
{
\begin{tabular}{ccccccccccc}
{} & {} &\multicolumn{3}{c}{Expansion}&\multicolumn{3}{c}{Rotation}&\multicolumn{3}{c}{Age}\\
{} & {} & $\kappa_X$ & $\kappa_Y$ & $\kappa_Z$ & $\omega_X$ & $\omega_Y$ & $\omega_Z$&${\kappa_X}^{-1}$&${\kappa_Y}^{-1}$&$\kappa_{\overline{XY}}^{-1}$\\
{Origin} & {Authors} & [$\rm{m\,s^{-1}\,pc^{-1}}$] & [$\rm{m\,s^{-1}\,pc^{-1}}$] & [$\rm{m\,s^{-1}\,pc^{-1}}$] 
& [$\rm{m\,s^{-1}\,pc^{-1}}$] & [$\rm{m\,s^{-1}\,pc^{-1}}$] & [$\rm{m\,s^{-1}\,pc^{-1}}$] & [Myr] & [Myr] & [Myr]\\
\hline 
\hline 
\multirow{2}{*}{Literature}&\citet{2014MNRAS.445.2169M}& $39\pm24$& $52\pm19$& $-31\pm44$&&& &$26_{-10}^{+41}$&$19_{-5}^{+11}$&$21_{-5}^{+10}$\\
                           &\citet{2020AA...642A.179M} & $57\pm6$ & $33\pm8$ & $-2\pm2$  &&& &$17\pm2$&$29\pm4$&$20\pm4$\\
\hline
\multirow{3}{*}{This work}&\citet{2019MNRAS.489.3625C} & $60\pm5$ & $34\pm11$& $0\pm17$  &$12\pm8$&$-5\pm8$  &$-6\pm7$ &$16.2_{-1.2}^{+1.5}$&$28.8_{-6.8}^{+13.5}$&$17.5_{-1.3}^{+1.6}$\\
                           &\citet{2020AA...642A.179M} & $59\pm5$ & $29\pm9$ & $-32\pm20$&$9\pm8$ &$-11\pm11$&$0\pm5$  &$16.6_{-1.2}^{+1.6}$&$33.6_{-6.9}^{+15.4}$&$19.0_{-1.5}^{+1.8}$\\
                           &\citet{2023ApJ...946....6C}& $51\pm4$ & $16\pm9$ & $-22\pm18$&$14\pm8$&$3\pm8$   &$-6\pm4$ &$19.3_{-1.4}^{+1.9}$&$61.1_{-22.5}^{+61.6}$&$22.0_{-1.8}^{+2.1}$\\
\hline 
\end{tabular}%
}
\end{table*}

We fit the linear velocity model (see Sect. \ref{method:linear_models}) to the $\beta$-Pic members from \citet{2020AA...642A.179M}, \citet{2019MNRAS.489.3625C}, and \citet{2023ApJ...946....6C}. However, in these two membership lists, we increased the scaling factor of the sky uncertainties to $10^7$ due to convergence failures with the default value of $10^6$ (see Sect. \ref{method:likelihood}). The most likely reason for these convergence issues is the presence of outliers and the lack of flexibility of the linear velocity model (see Sect. \ref{validation:discussion}). 

Table \ref{table:betapic} shows the result of our inference with the linear velocity model (last three rows) together with the literature values of the Galactic components of expansion reported by \citet{2014MNRAS.445.2169M} and \citet{2020AA...642A.179M}. As can be observed, our inferred values are comparable to those of the previous authors, except for the $\kappa_Z$ components reported by \citet{2020AA...642A.179M} and inferred here on the members by \citet{2019MNRAS.489.3625C}. Regarding rotation, we observe only mild evidence of it, with significance values in the order of $\lesssim 1\sigma$.

The expansion rates can be used to compute the expansion age of the system, $\tau$, using the relation $\tau\!=\!\gamma^{-1}\kappa^{-1}$, where $\gamma\!=\!1.022712165 \rm{s\,pc\, km^{-1}\, Myr^{-1}}$; see for example \citet{2014MNRAS.445.2169M} and \citet{2020AA...642A.179M}. Following \citet{2014MNRAS.445.2169M}, we discard the $\kappa_z$ component and compute expansion ages not only from $\kappa_X$ and $\kappa_Y$ independently, but also from their weighted average. The last three columns of Table \ref{table:betapic} show the expansion ages computed with this method as reported in the literature (first two rows) and obtained here (last three rows). As can be observed, the expansion age computed from the weighted average of $\kappa_X$ and $\kappa_Y$ results in values that agree with the dynamical trace-forward age of $17.8\!\pm\!1.2$ Myr by \citet{2019MNRAS.489.3625C} and the dynamical trace-back ages of $18.5_{-2.4}^{+2.0}$ Myr by \citet{2020AA...642A.179M} and $20.4\!\pm\!2.5$ Myr by \citet{2023ApJ...946....6C}. Combining the expansion ages inferred here (last column of Table \ref{table:betapic}) through a weighted average, we obtain a dynamical expansion age of $19.1\pm1.0$ Myr.

The excellent agreement and the similar or even better uncertainties of our expansion ages shows that our comprehensive statistical model combined with a simple inversion of the expansion rate gives age estimates that are as accurate and precise as the literature trace-back and trace-forward values. Nonetheless, we notice that this inversion method is highly sensitive to the uncertainties of the expansion rate, in a similar way that distance determination is sensitive to the parallax uncertainty when distance is computed as the inverse of the parallax. Therefore, in the presence of large uncertainties in the expansion rates, we recommend the use of more sophisticated methods, such as dynamical trace-back \citep[e.g.][]{2020AA...642A.179M}.

Summarising, \textit{Kalkayotl} was able to recover the source-level and population-level parameters of $\beta$-Pic reported in the literature with good accuracy and precision. The observed discrepancies in the population level parameters can be explained by the different statistical models assumed in the literature. Furthermore, we detect expansion at 12$\sigma$, 3$\sigma$, and 1$\sigma$ levels in the X, Y, and Z, directions, respectively, which results in expansion ages which are as precise and accurate as the literature ones that use more sophisticated methodologies. Moreover, we detect, for the first time, some hints of rotation in this system at the 1$\sigma$ level. Future work will be needed to confirm the existence of this rotation signal.

\subsection{The Hyades open cluster}
\label{hyades}

The Hyades open cluster is the closest to the Sun and has hundreds of members \citep[710 members at $47\pm0.2$ pc, $640_{-49}^{+67}$ Myr,][]{2019AA...623A..35L}, which makes it an excellent benchmark for kinematic analyses. Several works have analysed the internal kinematics of this cluster \citep[e.g.][]{2024A&A...687A..89J,2024ApJ...963..153H,2020MNRAS.498.1920O,2019MNRAS.483.5026L,2013ARep...57...52V,2000A&A...356.1119L,1998A&A...331...81P,1988AJ.....96..198G,1975AJ.....80..379H,1967PASP...79..156W} and recently its elongated tidal tails \citep[e.g.][]{2019AA...623A..35L,2019A&A...621L...3M,2019A&A...621L...2R}. Despite being thoroughly analysed in the literature, its internal rotation is still under debate. \citet{1967PASP...79..156W} studied the influence that internal motions such as contraction and rotation had on distance determinations made with the convergent point method. Using a combined analysis with proper motions, parallaxes, and radial velocities, he measured a contraction of $-13\pm15\,\rm{m\,s^{-1}pc^{-1}}$ and a total rotation of $|\omega|=30\pm40\,\rm{m\,s^{-1}pc^{-1}}$, thus concluding that their effect was negligible, in the case of rotation, and could amount up to 0.14 mag in the distance modulus, in the case of contraction. \cite{1975AJ.....80..379H} also studied the impact that rotation may have on distance determinations made with the convergent point method but now employing numerical simulations. He generated Hyades-like clusters with varying degrees of rotation values and orientations that were transformed to the observed space of proper motions, parallaxes, and radial velocities where he applied the convergent point method. He concluded that when $|\omega|\!=\!50\,\rm{m\,s^{-1}pc^{-1}}$ no significant effect was observed in the stellar motions, the convergent point solution, or the cluster distance. On the other hand, in the case of $|\omega|\!=\!500\,\rm{m\,s^{-1}pc^{-1}}$ the effects on the proper motions and radial velocities were so large that they would be easily spotted if the Hyades possessed such rotation.  

\cite{1988AJ.....96..198G} used precise radial velocities to determine a convergent point solution that was later used in combination with the full astrometric and radial velocity data to infer a dynamical model that included elongation and cluster rotation among other parameters. They inferred a distance of $45.4\pm2.1$ pc, a velocity dispersion of 230 $\rm{m\,s^{-1}}$, and a rotation gradient of $1.3\pm0.53\,\rm{km\,s^{-1}\,radians^{-1}}(28\pm11\,\rm{m\,s^{-1}\,pc^{-1}})$. Interestingly, they concluded that the cluster centre rotates faster than the outside but they were unable to determine the sense of the rotation (i.e. left-handed or right-handed).

The \textit{Hipparcos} data of this cluster was used by \citet{1998A&A...331...81P}, \cite{2000A&A...356.1119L} and \citet{2013ARep...57...52V}, among others, to analyse the internal kinematics of the cluster. \citet{1998A&A...331...81P} concluded that the cluster members move with parallel space motions and an internal velocity dispersion of $0.3\,\rm{km\,s^{-1}}$. They analysed the residuals of the space motions (see their Fig. 9) and concluded that although suggestive of a pattern of rotation or shearing motion, they were fully compatible within $3\sigma$ of the cluster motion. \cite{2000A&A...356.1119L} studied the cluster kinematics with a reduced version of their linear velocity model (see Sect. \ref{introduction}) that does not incorporate rotation. \citet{2013ARep...57...52V} applied a geometric model and measured, in the plane of the sky, a total internal rotation of $40\pm30\,\rm{m\,s^{-1}pc^{-1}}$. These authors also concluded that the rotation axis does not lie in the plane of the sky, and thus, that the study of the cluster's rotation demands not only proper motions and parallaxes but also radial velocities. 

\citet{2019MNRAS.483.5026L} use the spatial correlation of the differences between spectroscopic high-precision (160 $\rm{m\,s^{-1}}$) radial velocities with those predicted by Eq. 2 of \citet{2002A&A...381..446M}, and measure the cluster's velocity gradient, in the right ascension coordinate. This gradient translates into a rotation signal of $42.3\pm4.0\,\rm{m\,s^{-1}pc^{-1}}$. 

\citet{2020MNRAS.498.1920O} use an improved version of the general model of \cite{2000A&A...356.1119L} and apply it to the \textit{Gaia} Data Release 2 \citep[DR2;][]{2018A&A...616A...1G} of the Hyades members from  \citet{2019A&A...621L...3M} and \citet{2019A&A...621L...2R}. They find no evidence of rotation.

\citet{2024ApJ...963..153H} used \textit{Gaia} DR3 data from the Hyades members by \citet{2019AA...623A..35L} to analyse its internal rotation. They found a mean rotational velocity of $90\pm30\,\rm{m\,s^{-1}}$ within the cluster tidal radius of 9 pc. To compare with the common units of rotation, we use instead the peak value of their rotational curve (see their Table 8 and Fig. 14), which is located at 3 pc with a value of $0.61\pm0.06\,\rm{km\,s^{-1}}$ which corresponds to $203\pm20\,\rm{m\,s^{-1}\,pc^{-1}}$.

\citet{2024A&A...687A..89J} used \textit{Gaia} DR3 data from the members by \cite{2023A&A...673A.114H} to study the internal kinematic the Hyades. However, they do not find evidence for rotation.

Given that our Gaussian linear velocity model is similar to that of \citet{2020MNRAS.498.1920O}, we compare our results of rotation and expansion with those of the previous authors using their list of members with data from both the second and third releases of \textit{Gaia}. In particular, we work with the list of members from the cluster core, which is expected to be less subject to Galactic shear forces. In addition, we expect the core list of members to have fewer contaminants than that of the tails, given that the tails are more confused with the Galactic field than the high-contrast and dense region of the core. We notice that \citet{2020MNRAS.498.1920O} report their values in the ICRS reference frame, and thus, when comparing with those authors, we adopted this same frame.

First, we work with the \textit{Gaia} DR2 data (i.e. the same astrometry of \citealt{2020MNRAS.498.1920O}) to which we applied our decontamination algorithm (see Sect. \ref{method:FGMM}). We identified 13 outliers representing a fraction of members of $0.968\pm0.008$. This fraction overlaps at the 95\% HDI with the $0.953\pm0.013$ value reported by \citet{2020MNRAS.498.1920O}. Then, we applied the Gaussian linear velocity model (see Sect. \ref{method:linear_models}) to the remaining 387 core members. The inferred posterior distributions of our linear model parameters are shown and compared to those of \citet{2020MNRAS.498.1920O} in Table \ref{table:hyades_grp}. We notice that the model of those authors is only for the velocity field; thus it lacks the $loc$ and $std$ parameters of the X, Y, and Z dimensions.

\begin{table}[ht!]
\caption{Parameters of the Hyades open cluster as reported by \citet{2020MNRAS.498.1920O} and inferred here with data from \textit{Gaia} DR2 and DR3.}
\label{table:hyades_grp}
\centering
\resizebox{\columnwidth}{!}
{
\begin{tabular}{llrrr}
\toprule
 &  & Oh+2020 & \textit{Gaia} DR2 & \textit{Gaia} DR3 \\
Parameter & Units &  &  &  \\
\midrule
loc[X] & [pc] & - & $17.14\pm0.15$ & $17.11\pm0.15$ \\
loc[Y] & [pc] & - & $41.14\pm0.17$ & $41.06\pm0.17$ \\
loc[Z] & [pc] & - & $13.63\pm0.17$ & $13.60\pm0.18$ \\
loc[U] & $\rm{[km\, s^{-1}]}$ & $-6.09\pm0.03$ & $-6.05\pm0.03$ & $-6.01\pm0.03$ \\
loc[V] & $\rm{[km\, s^{-1}]}$ & $45.63\pm0.05$ & $45.62\pm0.05$ & $45.69\pm0.06$ \\
loc[W] & $\rm{[km\, s^{-1}]}$ & $5.52\pm0.03$ & $5.53\pm0.03$ & $5.58\pm0.03$ \\
std[X] & [pc] & - & $2.92\pm0.10$ & $2.91\pm0.11$ \\
std[Y] & [pc] & - & $3.33\pm0.12$ & $3.33\pm0.12$ \\
std[Z] & [pc] & - & $3.46\pm0.13$ & $3.46\pm0.13$ \\
std[U] & $\rm{[km\, s^{-1}]}$ & $0.44\pm0.07$ & $0.44\pm0.03$ & $0.50\pm0.03$ \\
std[V] & $\rm{[km\, s^{-1}]}$ & $0.38\pm0.02$ & $0.44\pm0.04$ & $0.66\pm0.05$ \\
std[W] & $\rm{[km\, s^{-1}]}$ & $0.37\pm0.06$ & $0.37\pm0.02$ & $0.39\pm0.02$ \\
$\rho_{UV}$ &  & $-0.15\pm0.37$ & $0.20\pm0.11$ & $0.38\pm0.07$ \\
$\rho_{UW}$ &  & $-0.01\pm0.30$ & $0.09\pm0.08$ & $0.19\pm0.07$ \\
$\rho_{VW}$ &  & $-0.17\pm0.17$ & $0.06\pm0.12$ & $0.29\pm0.08$ \\
$||\kappa||$ & $\rm{[m\,s^{-1}\,pc^{-1}]}$ & $-6.50\pm6.42$ & $-6.38\pm7.26$ & $-1.72\pm8.06$ \\
$\omega_x$ & $\rm{[m\,s^{-1}\,pc^{-1}]}$ & $3.27\pm5.51$ & $-4.99\pm9.76$ & $-3.60\pm10.31$ \\
$\omega_y$ & $\rm{[m\,s^{-1}\,pc^{-1}]}$ & $2.24\pm9.78$ & $1.42\pm6.58$ & $-0.09\pm7.05$ \\
$\omega_z$ & $\rm{[m\,s^{-1}\,pc^{-1}]}$ & $-4.44\pm8.71$ & $6.49\pm10.11$ & $29.02\pm11.79$ \\
$w_1$ & $\rm{[m\,s^{-1}\,pc^{-1}]}$ & $1.45\pm5.46$ & $-7.04\pm9.69$ & $2.29\pm10.46$ \\
$w_2$ & $\rm{[m\,s^{-1}\,pc^{-1}]}$ & $-6.59\pm10.07$ & $6.49\pm6.33$ & $1.55\pm6.76$ \\
$w_3$ & $\rm{[m\,s^{-1}\,pc^{-1}]}$ & $1.66\pm8.70$ & $-14.46\pm10.94$ & $-33.92\pm12.70$ \\
$w_4$ & $\rm{[m\,s^{-1}\,pc^{-1}]}$ & $-11.19\pm15.60$ & $-13.56\pm10.79$ & $-30.24\pm11.95$ \\
$w_5$ & $\rm{[m\,s^{-1}\,pc^{-1}]}$ & $10.64\pm6.32$ & $-9.48\pm16.32$ & $17.57\pm18.49$ \\
\bottomrule
\end{tabular}
}
\end{table}

As can be observed from Table \ref{table:hyades_grp}, in the same data as \cite{2020MNRAS.498.1920O}, this is column \textit{Gaia} DR2, we recover the location parameters of the 3D velocities ($U,V,W$) with similar precision, while their standard deviations ($std[U]$ and $std[W]$) and correlations with better ones. Only the dispersion in the $V$ coordinate is recovered with lower precision but it is, nonetheless, compatible with the uncertainties as the rest of the parameters. Furthermore, all of the linear velocity parameters (i.e. $|\vec{\kappa}|,\vec{\omega}$, and $\vec{w}$) are consistent within the uncertainties with those of \cite{2020MNRAS.498.1920O} although with varying precision. 

Then, we proceed to analyse the \textit{Gaia} DR3 data of the same 387 core members with the Gaussian linear velocity model. The results of this inference are also shown in Table \ref{table:hyades_grp}, under the column \textit{Gaia} DR3. These results show that there is only a mild improvement in the precision of the correlation parameters ($\rho_{UV}$ and $\rho_{VW}$). However, the most relevant result is the detection, at the $2\sigma$ level, of rotation in $\omega_z$. 

To compare with other works from the literature, we also inferred the parameters of the linear velocity model in the Galactic reference frame. Unfortunately, from the literature works that report the cluster parameters in this frame, we found that only \cite{2019AA...623A..35L} provides the associated uncertainties. These authors report the cluster location parameters for sources located within two radii: 10 pc and 20 pc. We use the value 10 pc because it coincides with the selection criteria of \citealt{2020MNRAS.498.1920O}, from which we take the list of members for the cluster core. In Table \ref{table:hyades_location}, we compare the location parameters reported by \cite{2019AA...623A..35L} with those inferred here using the \textit{Gaia} DR3 data and the linear velocity model. As can be observed, the parameters show an overall good agreement despite being inferred from different lists of members and data sets.  

\begin{table}[ht!]
\caption{Location parameters of the Hyades cluster in the Galactic frame.}
\label{table:hyades_location}
\centering
\begin{tabular}{cccc}
Parameter & Units & \citet{2019AA...623A..35L} & This work \\
{} & {} & \textit{Gaia} DR2 & \textit{Gaia} DR3 \\
\hline 
\hline 
$X$ & [pc]                & $-43.83\pm0.18$ & $-43.39\pm0.18$ \\
$Y$ & [pc]                & $0.42\pm0.11$   & $0.35\pm0.17$   \\
$Z$ & [pc]                & $-17.05\pm0.09$ & $-16.78\pm0.15$ \\
$U$ & [$\rm{km\,s^{-1}}$] & $-42.14\pm0.11$ & $-42.28\pm0.06$ \\
$V$ & [$\rm{km\,s^{-1}}$] & $-19.26\pm0.04$ & $-19.13\pm0.02$ \\
$W$ & [$\rm{km\,s^{-1}}$] & $-1.12\pm0.05$  & $-1.29\pm0.03$  \\
\hline 
\end{tabular}
\end{table}

In the Galactic reference frame the rotation signal remains at the $2\sigma$ level; however it is now in $\omega_x$ alone, with a value of $\vec{\omega}\!=\![-13.61\pm6.21,19.87\pm11.82,15.93\pm10.30]\,\rm{m\,s^{-1}\,pc^{-1}}$. Unfortunately, this rotation signal cannot be directly compared to the values reported in the literature given that the majority of these were obtained by analysing the velocity gradients in the plane of the sky, whereas our model returns a vector of gradients in the 3D space. Nonetheless, the norm of the rotation vector we infer here is perfectly compatible, within the 95\% HDI, with the total rotation values of the literature works (see Table \ref{table:hyades_rotation}), except with the recent value of \cite{2024ApJ...963..153H}, which exceeds the literature values  by more than five times. We notice that if the Hyades has a rotation signal $\sim\!200\,\rm{m\,s^{-1}\,pc^{-1}}$, then our method would detect it on the \textit{Gaia} DR3 data with a significance level higher than 5$\sigma$ (see Fig. \ref{figure:snr}).

\begin{table}[ht!]
\caption{Literature values of the Hyades rotational signal.}
\label{table:hyades_rotation}
\centering
\begin{tabular}{cc}
Authors & $|\vec{\omega}|$ \\
{} & [$\rm{m\,s^{-1}\,pc^{-1}}$] \\
\hline 
\hline 
\citet{1967PASP...79..156W} & $30\pm40$ \\  
\citet{1988AJ.....96..198G} & $28\pm11$ \\ 
\citet{2013ARep...57...52V} & $40\pm30$\\
\citet{2019MNRAS.483.5026L} & $42.3\pm4.0$ \\
\citet{2024ApJ...963..153H} & $203\pm20$ \\
This work & $32.1\pm11.0$ \\
\hline 
\end{tabular}
\end{table}

Summarising, the Gaussian linear velocity model implemented here recovers, with excellent accuracy and slightly better precision, the cluster's parameters reported by \citealt{2020MNRAS.498.1920O} in the ICRS frame and those of \cite{2019AA...623A..35L} in the Galactic frame. Furthermore, when applied over the \textit{Gaia} DR3 data, it unravels, at the $2\sigma$ level, the rotational signal of this cluster and confirms most of the rotation values from the literature. More precise and complete radial velocity measures are needed to claim the discovery of the Hyades rotation at a higher significance level.

\subsection{The Praesepe open cluster}
\label{praesepe}

\begin{table*}[ht!]
\caption{Parameters of the Praesepe open cluster as inferred here with \textit{Gaia} DR3 data from the samples of members (original and restricted to the tidal radius) reported by \cite{2022ApJ...938..100H}, \cite{2023A&A...673A.128G}, and \citet{2024A&A...687A..89J}.}
\label{table:praesepe_grp}
\centering
\resizebox{\textwidth}{!}
{
\begin{tabular}{llrrrrrr}
\toprule
 & Case & \multicolumn{3}{c}{Original} & \multicolumn{3}{c}{Cleaned \& <$R_{tidal}$} \\
 & Origin & Hao+2022 & GG+2023 & Jadhav+2024 & Hao+2022 & GG+2023 & Jadhav+2024 \\
Parameter & Units &  &  &  &  &  &  \\
\midrule
loc[X] & [pc] & $-140.45\pm0.53$ & $-139.63\pm0.34$ & $-140.00\pm0.31$ & $-139.34\pm0.39$ & $-139.68\pm0.31$ & $-139.94\pm0.30$ \\
loc[Y] & [pc] & $-68.45\pm0.45$ & $-67.97\pm0.18$ & $-68.03\pm0.18$ & $-67.76\pm0.26$ & $-68.01\pm0.17$ & $-68.01\pm0.17$ \\
loc[Z] & [pc] & $98.43\pm0.46$ & $98.59\pm0.24$ & $98.62\pm0.23$ & $98.40\pm0.31$ & $98.65\pm0.23$ & $98.59\pm0.22$ \\
loc[U] & $\rm{[km\, s^{-1}]}$ & $-42.65\pm0.08$ & $-42.52\pm0.10$ & $-42.62\pm0.09$ & $-42.66\pm0.07$ & $-42.66\pm0.07$ & $-42.67\pm0.07$ \\
loc[V] & $\rm{[km\, s^{-1}]}$ & $-20.32\pm0.07$ & $-20.31\pm0.05$ & $-20.35\pm0.05$ & $-20.27\pm0.06$ & $-20.37\pm0.04$ & $-20.37\pm0.04$ \\
loc[W] & $\rm{[km\, s^{-1}]}$ & $-9.60\pm0.12$ & $-9.55\pm0.09$ & $-9.57\pm0.08$ & $-9.42\pm0.08$ & $-9.51\pm0.08$ & $-9.52\pm0.07$ \\
std[X] & [pc] & $5.40\pm0.32$ & $3.79\pm0.13$ & $2.90\pm0.10$ & $3.34\pm0.21$ & $3.20\pm0.11$ & $2.90\pm0.10$ \\
std[Y] & [pc] & $5.41\pm0.30$ & $2.91\pm0.09$ & $2.81\pm0.08$ & $2.83\pm0.17$ & $2.77\pm0.08$ & $2.81\pm0.08$ \\
std[Z] & [pc] & $5.22\pm0.30$ & $2.89\pm0.09$ & $2.69\pm0.08$ & $2.94\pm0.18$ & $2.63\pm0.08$ & $2.68\pm0.08$ \\
std[U] & $\rm{[km\, s^{-1}]}$ & $0.82\pm0.06$ & $1.28\pm0.09$ & $1.05\pm0.08$ & $0.60\pm0.06$ & $0.80\pm0.06$ & $0.76\pm0.06$ \\
std[V] & $\rm{[km\, s^{-1}]}$ & $0.72\pm0.05$ & $0.89\pm0.04$ & $0.65\pm0.03$ & $0.68\pm0.04$ & $0.62\pm0.03$ & $0.56\pm0.03$ \\
std[W] & $\rm{[km\, s^{-1}]}$ & $1.05\pm0.07$ & $1.09\pm0.07$ & $0.79\pm0.05$ & $0.60\pm0.05$ & $0.69\pm0.04$ & $0.63\pm0.04$ \\
$||\kappa||$ & $\rm{[m\,s^{-1}\,pc^{-1}]}$ & $20.04\pm8.18$ & $-5.22\pm11.11$ & $7.56\pm12.77$ & $1.16\pm11.37$ & $-18.24\pm9.90$ & $-2.27\pm10.55$ \\
$\kappa_x$ & $\rm{[m\,s^{-1}\,pc^{-1}]}$ & $15.04\pm15.71$ & $15.83\pm27.94$ & $1.11\pm27.18$ & $10.08\pm18.91$ & $3.33\pm21.60$ & $3.18\pm22.95$ \\
$\kappa_y$ & $\rm{[m\,s^{-1}\,pc^{-1}]}$ & $60.04\pm11.15$ & $3.19\pm18.54$ & $0.09\pm15.15$ & $20.51\pm21.08$ & $-22.95\pm13.82$ & $-4.37\pm13.26$ \\
$\kappa_z$ & $\rm{[m\,s^{-1}\,pc^{-1}]}$ & $-14.96\pm17.16$ & $-34.68\pm26.10$ & $21.48\pm22.25$ & $-27.11\pm19.91$ & $-35.11\pm19.20$ & $-5.62\pm18.56$ \\
$\omega_x$ & $\rm{[m\,s^{-1}\,pc^{-1}]}$ & $-49.70\pm10.13$ & $-7.25\pm16.46$ & $-1.97\pm13.10$ & $2.02\pm14.54$ & $-6.08\pm11.82$ & $-8.87\pm11.48$ \\
$\omega_y$ & $\rm{[m\,s^{-1}\,pc^{-1}]}$ & $49.67\pm12.46$ & $53.06\pm22.40$ & $-17.57\pm18.53$ & $5.09\pm14.46$ & $5.94\pm15.74$ & $3.29\pm15.04$ \\
$\omega_z$ & $\rm{[m\,s^{-1}\,pc^{-1}]}$ & $11.35\pm9.35$ & $8.69\pm20.14$ & $-9.55\pm16.75$ & $-7.37\pm14.85$ & $1.05\pm14.18$ & $-4.92\pm14.18$ \\
$w_1$ & $\rm{[m\,s^{-1}\,pc^{-1}]}$ & $26.13\pm10.10$ & $-18.03\pm15.36$ & $-7.75\pm13.63$ & $-3.58\pm14.77$ & $6.39\pm11.52$ & $5.28\pm11.57$ \\
$w_2$ & $\rm{[m\,s^{-1}\,pc^{-1}]}$ & $30.61\pm11.68$ & $3.43\pm17.60$ & $-1.29\pm18.54$ & $17.28\pm13.80$ & $17.36\pm14.36$ & $18.79\pm15.27$ \\
$w_3$ & $\rm{[m\,s^{-1}\,pc^{-1}]}$ & $4.43\pm9.01$ & $4.68\pm16.76$ & $-2.13\pm15.96$ & $3.58\pm14.29$ & $-7.43\pm13.07$ & $-6.62\pm13.30$ \\
$w_4$ & $\rm{[m\,s^{-1}\,pc^{-1}]}$ & $15.04\pm15.71$ & $15.83\pm27.94$ & $1.11\pm27.18$ & $10.08\pm18.91$ & $3.33\pm21.60$ & $3.18\pm22.95$ \\
$w_5$ & $\rm{[m\,s^{-1}\,pc^{-1}]}$ & $60.04\pm11.15$ & $3.19\pm18.54$ & $0.09\pm15.15$ & $20.51\pm21.08$ & $-22.95\pm13.82$ & $-4.37\pm13.26$ \\
\bottomrule
\end{tabular}
}
\end{table*}
The \object{Praesepe} open cluster \citep[$187.4\pm3.9$ pc, $673_{-39}^{+55}$ Myr, and more than 2000 members,][]{2019A&A...628A..66L}, has recently been at the centre of a controversy about its rotational velocity. The works by \cite{2024A&A...687A..89J}, \citet{2020AN....341..638L}, \citet{2022ApJ...938..100H}, and \citet{2023A&A...673A.128G} have analysed this cluster with independent data and methodologies, finding contradictory results. \citet{2020AN....341..638L} find a possible rotation velocity of $400\,\rm{m\,s^{-1}}$ in the periphery of the cluster. \citet{2022ApJ...938..100H} find a clear rotation signal with a velocity of $200\pm50\,\rm{m\,s^{-1}}$ within their cluster's tidal radius of 10 pc. \citet{2023A&A...673A.128G} do not find rotation and suggest that when found, it results from uncorrected projection effects (see their Appendix B and Fig. B.1). Finally, based only on radial velocity data, \citet{2024A&A...687A..89J} find a non-significant (0.5$\sigma$) rotational signal of $9\pm18\,\rm{cy\,Gyr^{-1}}$, equivalent to $9.2\pm18.4\,\rm{m\,s^{-1}\,pc{-1}}$.

Here, we explore the internal kinematics of Praesepe with the Gaussian linear velocity model applied to the \textit{Gaia} DR3 data of the list of members utilised by \cite{2022ApJ...938..100H}, \cite{2023A&A...673A.128G} and \citet{2024A&A...687A..89J}. Given the controversial results of these works, we run the Gaussian linear velocity model directly on their list of members as well as in the cleaned subsamples resulting after applying our decontamination algorithm and restricting the analysis to sources within the tidal radius of 10.7 pc reported by  \cite{2019A&A...628A..66L}.  After applying our decontamination algorithm with a field scale of 20 pc in position and 5 $\rm{km\, s^{-1}}$ in velocity (and further restricting to the cluster's  tidal radius), we remove 12 (10), 30 (7), and 3 (0) sources from  the membership lists by \cite{2022ApJ...938..100H}, \cite{2023A&A...673A.128G}, and \cite{2024A&A...687A..89J}, respectively. We notice that all original members by \citet{2024A&A...687A..89J} were contained within the cluster's tidal radius. Table \ref{table:praesepe_grp} shows summaries of the posterior distributions from the parameters of the Gaussian linear velocity model as inferred in all previous cases.

In the original samples, all the location parameters agree with themselves within the uncertainties. However, the standard deviations in the positions X, Y, and Z are discrepant, with those inferred from the members by \citet{2022ApJ...938..100H} being almost two times larger than those of the other cases. On the contrary, the standard deviations in the velocities U, V, and W are similar in the three cases. On the other hand, in the samples that were cleaned and restricted to the tidal radius, we observe that the location parameters continue to be in agreement with all previous cases and the velocity dispersion also show similar although more consistent values <1 $\rm{km\,s^{-1}}$. However, contrary to what was observed in the original samples, the standard deviations in positions X, Y, and Z are all now similar, with values $\lesssim$3 pc.

The internal motions in the original samples show the following results. First, expansion is only detected at $2\sigma$ level in the sample from \cite{2022ApJ...938..100H}, the other two samples show spurious detections (<1$\sigma$). Second, rotation in $\omega_x$ and $\omega_y$ is detected at 4$\sigma$ on the sample by \cite{2022ApJ...938..100H} and at 2$\sigma$ in $\omega_y$ on the sample by \cite{2023A&A...673A.128G}. The sample by \cite{2024A&A...687A..89J} shows a spurious detection ($\sim$1$\sigma$) of rotation in $\omega_y$ alone although with opposite sign.

The results of the inference on the samples that were clean and restricted to the tidal radius indicate no significant detections of rotation and only some hints of contraction along the Z component. The rotational signal we found in all cases shows no-significant rotations in the three components, which is consistent with the recent findings by \cite{2024A&A...687A..89J} along the line of sight.

We conclude that the signals of expansion and rotation, and rotation found on the original lists of members used by \cite{2022ApJ...938..100H} and \cite{2023A&A...673A.128G}, respectively, result from the cluster members beyond the tidal radius. We found no conclusive evidence for the rotation of the cluster's central region within its tidal radius. Thus, we confirm that Praesepe's rotational signal reported in the literature comes from its periphery (i.e. halo or tails). Moreover, our analysis shows that the apparent contradiction  in the literature results roots in the contaminants present in the various membership lists.

\section{Conclusions}
\label{conclusions}

In this work, we introduce, develop, and validate a free and open-source code for the inference of the phase-space parameters of star-forming regions, stellar associations, and open clusters. This code is the multidimensional extension of an already published methodology that enables its users to simultaneously infer global-level and source-level parameters of LNSS based on \textit{Gaia} data. 

This new code offers several methodological improvements with respect to similar ones from the literature, among which we mention the simultaneous modelling of positions, velocities, and their correlations, the treatment of parallax and proper motions angular correlations, the flexibility of its statistical models, the treatment of sources with missing values, particularly radial velocities, decontamination algorithms to clean up the input list of members, and the modelling of kinematic substructures. Furthermore, the comprehensive modelling enables the improvement of radial velocities and parallaxes through astrometric radial velocities and kinematically improved parallaxes \citep[see][]{2000A&A...356.1119L}.

Our validation with real data shows that the linear velocity model that we present here offers similar accuracy and slightly better precision than the one implemented by \cite{2020MNRAS.498.1920O}. Nonetheless, \textit{Kalkayotl} includes an improved version of this linear velocity model with Gaussian and Student-T distributions both in the spatial and velocity components. Moreover, thanks to the efficient Hamiltonian Monte Carlo sampling of the full posterior distribution of the improved \cite{2000A&A...356.1119L} linear velocity model our methodology can provide objective detectability criteria for internal kinematic patterns. Thus, the methodology present here provides the community with a free and open source code that delivers objective and reproducible criteria for the detection of patterns of expansion, contraction, and rotation. 

We applied our newly developed methodology on benchmark stellar systems where extensive kinematic information is available in the literature. Our results show good accuracy and precision, resulting in agreement with the parameter values reported in the literature. We notice that the results of the internal kinematic patterns are sensitive to contaminants or outliers in the input list of members. After applying our decontamination methodology, we found the following results. 

In the $\beta$-Pic stellar association, \textit{Kalkayotl} joint velocity model recovers the source- and group-level parameters from the literature with excellent accuracy and slightly better precision. The Gaussian linear velocity models detects expansion at the $2\sigma$ level. By simple inversion, these expansion values translate into ages that are as accurate and precise as those from the literature. 

In the Hyades open cluster, the Gaussian linear velocity model recovers the literature parameter values with similar accuracy and slightly better precision. Furthermore, it detects rotation at the $2\sigma$ level in the $\omega_z$ component, with a total rotational magnitude compatible with the literature values.

In the Praesepe open cluster, we find no evidence of rotation within the cluster tidal radius but only in its outskirts. Thus, we conclude that the rotation and expansion signals reported in the literature come from the periphery of the cluster.

Although the mechanisms producing the expansion and rotation of stellar systems are well understood \citep[see, for example, the introduction of ][]{2024A&A...687A..89J}, the relative contribution of these different mechanisms is still poorly characterised. To improve this characterisation, both numerical N-body simulations with varying initial conditions \citep{2024A&A...687A..89J} and observational surveys that minimise biases arising from contaminants, incomplete membership lists, and kinematic substructures are still needed. The robust statistical methodology that we present here offers ways to characterise and minimise these biases in LNSS up to 1.5 kpc.

Finally, we notice that although our methodology offers several advantages over those from the literature, it is still computationally expensive and requires further statistical modelling of the physical properties of stellar systems, including their spatial distribution \citep{2018A&A...612A..70O}, differential rotation \citep{2024A&A...687A..89J}, or the tidal tails of open clusters. Future steps will be taken to continue improving \textit{Kalkayotl}'s  models.

\begin{acknowledgements}
We thank the anonymous referee for providing comments that greatly improved the quality of this work.
JO acknowledge financial support from "Ayudas para contratos postdoctorales de investigación UNED 2021".
"La publicación es parte del proyecto
PID2022-142707NA-I00, financiado por MCIN/AEI/10.13039/501100011033/FEDER, UE".
A. Berihuete was also funded by TED2021-130216A-I00 (MCIN/AEI/10.13039/501100011033 and European Union NextGenerationEU/PRTR).
This research has received funding from the European Research Council (ERC) under the European Union’s Horizon 2020 research and innovation programme (grant agreement No 682903, P.I. H. Bouy), and from the French State in the framework of the "Investments for the future" Program, IdEx Bordeaux, reference ANR-10-IDEX-03-02 .
This work has made use of data from the European Space Agency (ESA) mission
{\it Gaia} (\url{https://www.cosmos.esa.int/gaia}), processed by the {\it Gaia}
Data Processing and Analysis Consortium (DPAC,
\url{https://www.cosmos.esa.int/web/gaia/dpac/consortium}). Funding for the DPAC
has been provided by national institutions, in particular the institutions
participating in the {\it Gaia} Multilateral Agreement.
We express our gratitude to Anthony Brown, Jos de Bruijne, the Gaia Project Scientist Support Team, and the Gaia Data Processing and Analysis Consortium (DPAC) for providing the \textit{PyGaia} code.
We thank the \textit{PyMC} team for making publicly available this probabilistic programming language.
This research has made use of the VizieR catalogue access tool, CDS, Strasbourg, France.

\end{acknowledgements}

\bibliographystyle{aa} 
\bibliography{Kalkayotl} 

\begin{appendix}
\section{Assumptions}
\label{appendix:assumptions}

\begin{assumption}
\label{assumption:normality}
Same as \citetalias{2020A&A...644A...7O}. Briefly, the \textit{Gaia} measurements are normally distributed.
\end{assumption}

\begin{assumption}
\label{assumption:zeropoints}
Same as \citetalias{2020A&A...644A...7O} but extended to all astrometry and radial velocities. Briefly, the \textit{Gaia} measurements have zero point values for all its observables and these are specified by the user.
\end{assumption}

\begin{assumption}
\label{assumption:angular_correlations}
Same as \citetalias{2020A&A...644A...7O} but extended to proper motions as well. Briefly, the \textit{Gaia} parallax and proper motions are spatially correlated in the sky. To correct for these spatial correlations, we now use Eqs. 24 and 25 of \cite{2021A&A...649A...2L}.
\end{assumption}

\begin{assumption}
\label{assumption:clean_members}
Same as \citetalias{2020A&A...644A...7O}. Briefly, the input list of members is not biased.
\end{assumption}

In the 3D and 6D models presented here, the Assumption 4 of \citetalias{2020A&A...644A...7O} is no longer needed.

\section{Time scalability}
\label{appendix:scalability}
\begin{figure}[ht!]
    \centering
     \includegraphics[width=\columnwidth,page=1]{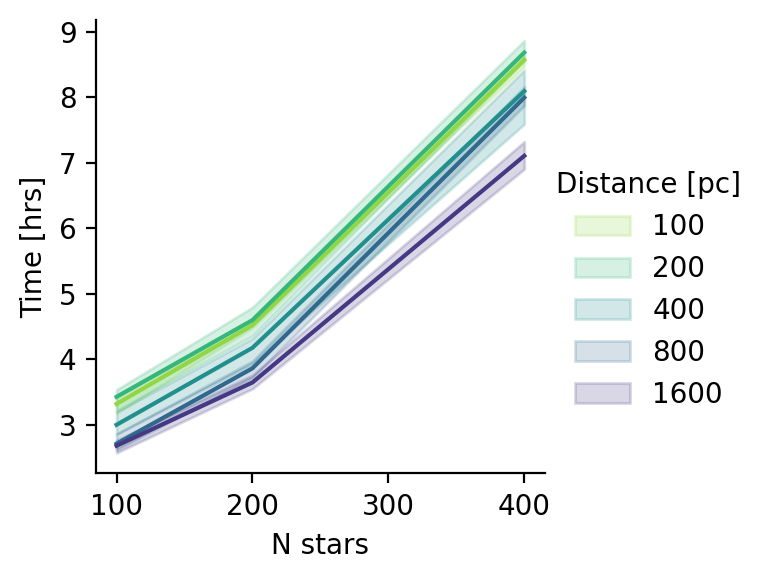}
     \caption{Time scalability of \textit{Kalkayotl}'s runs done with the \textit{numpyro} NUTS sampler as a function of the number of stars. The lines and shaded regions depict the mean and 2$\sigma$ percentiles, respectively, from the grid of stellar systems created using the Student-T family with the linear velocity model (see Sect. \ref{validation}).}
\label{figure:numpyro_time_scalability}
\end{figure}

\begin{figure}[ht!]
    \centering
     \includegraphics[width=\columnwidth,page=1]{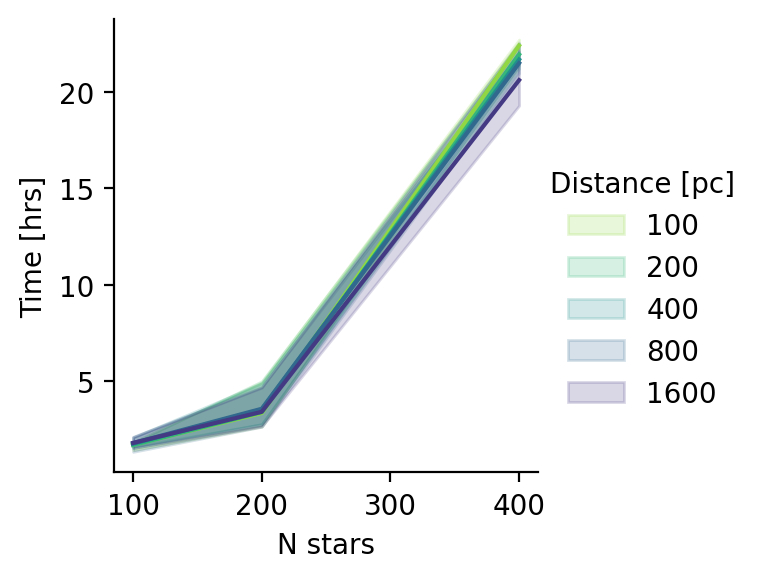}
     \caption{Time scalability of \textit{Kalkayotl}'s runs done with the native \textit{pymc} NUTS sampler as a function of the number of stars. The lines and shaded regions depict the mean and 2$\sigma$ percentiles, respectively, from the grid of stellar systems created using the GMM family with  velocity model (see Sect. \ref{validation}).}
\label{figure:pymc_time_scalability}
\end{figure}

In this Appendix, we aim at giving the users of \textit{Kalkayotl} a rough idea of the typical amount of time that an execution of the code may take. To this end, Figs. \ref{figure:numpyro_time_scalability} and \ref{figure:pymc_time_scalability} show the amount of time, in hours, as a function of the number of stars in the system for the Student-T family with linear velocity model and GMM family with joint velocity model, respectively. This graphs were computed from the simulations created in Sect. \ref{validation} and inferred with the NUTS \citep{2011arXiv1111.4246H} samplers \textit{numpyro} for the Student-T family and the \textit{pymc} native for the GMM family. The \textit{numpyro} implementation of the NUTS sampler (see PyMC documentation\footnote{\url{https://www.pymc.io/welcome.html}}) is faster than the native \textit{pymc} one. However, the current version of PyMC fails at sampling 6D GMM with the \textit{numpyro} sampler. We run the code using two parallel Markov chains in a machine with 2.1 GHz CPUs. The execution time takes into account not only the MCMC inference but all the auxiliary pre-processing of the data and post-processing of the chains. 

As expected and observed from Figs. \ref{figure:numpyro_time_scalability} and \ref{figure:pymc_time_scalability}, the execution time grows with the number of stars. In addition, the figures show that for a fixed number of stars, the execution time is smaller for systems at farther distances (colour code), with this difference being less pronounced in the GMM case. The smaller running time of faraway systems is explained by the large uncertainty of their observables, which simplifies the job of finding the initial parameter values and, thus, the subsequent sampling of the chains.

\end{appendix}

\end{document}